\documentclass[aps,prd,preprintnumbers,groupedaddress,nofootinbib,12pt]{revtex4}

\pdfoutput=1

\usepackage{graphicx}
\usepackage{latexsym}
\usepackage{amsfonts}
\usepackage{amssymb}
\usepackage{amsmath}
\usepackage{slashed}
\usepackage{array}
\usepackage{feynmp}
\usepackage{url}
\usepackage{multirow}
\newcommand{\gev}{{\rm GeV}}

\newcommand{\be}{\begin{equation}}
\newcommand{\ee}{\end{equation}}
\usepackage{accents}
\newlength{\dhatheight}

\def\bea{\begin{eqnarray}}
\def\eea{\end{eqnarray}}
\def\ltap{\ \raise.3ex\hbox{$<$\kern-.75em\lower1ex\hbox{$\sim$}}\ }
\def\gtap{\ \raise.3ex\hbox{$>$\kern-.75em\lower1ex\hbox{$\sim$}}\ }
\def\lsim{\ \raise.3ex\hbox{$<$\kern-.75em\lower1ex\hbox{$\sim$}}\ }
\def\gsim{\ \raise.3ex\hbox{$>$\kern-.75em\lower1ex\hbox{$\sim$}}\ }

\def\ie{{\it i.e.}}

\newcommand{\met}{\slashed {E}_{T}}
\usepackage{color}

\begin{document}
\title{Stops and $\met$: the shape of things to come}

\author{Daniele S.M.~Alves$^1$, Matthew R.~Buckley$^{1}$, Patrick~J.~Fox$^2$, Joseph D.~Lykken$^2$, and Chiu-Tien Yu$^{3,2}$}
\affiliation{$^1$Center for Particle Astrophysics, Fermi National Accelerator Laboratory, Batavia, IL 60510, USA}
\affiliation{$^2$Theoretical Physics Department, Fermi National Accelerator Laboratory, Batavia, IL 60510, USA}
\affiliation{$^3$ Department of Physics, University of Wisconsin, Madison, WI 53706 USA}

\preprint{FERMILAB-PUB-12-250-A-T}
\date{\today}

\begin{abstract}
LHC experiments have placed strong bounds on the production of supersymmetric colored particles (squarks and gluinos), under the assumption that all flavors of squarks are nearly degenerate. However, the current experimental constraints on stop squarks are much weaker, due to the smaller production cross section and difficult backgrounds. While light stops are motivated by naturalness arguments, it has been suggested that such particles become nearly impossible to detect near the limit where their mass is degenerate with the sum of the masses of their decay products. We show that this is not the case, and that searches based on missing transverse energy ($\met$) have significant reach for stop masses above $175$ GeV, even in the degenerate limit. We consider direct pair production of stops, decaying to invisible LSPs  and tops with either hadronic or semi-leptonic final states. Modest intrinsic differences in $\met$ are magnified by boosted kinematics and by shape analyses of $\met$ or suitably-chosen observables related to $\met$. For these observables we show that the distributions of the relevant backgrounds and signals are well-described by simple analytic functions, in the kinematic regime where signal is enhanced. Shape analyses of $\met$-related distributions will allow the LHC experiments to place significantly improved bounds on stop squarks, even in scenarios where the stop-LSP mass difference is degenerate with the top mass. Assuming 20~fb$^{-1}$ of luminosity at $\sqrt{s} = 8$~TeV, we conservatively estimate that experiments can exclude or discover degenerate stops with mass as large as $\sim 360$~GeV and 560 GeV for massless LSPs.
\end{abstract}

\maketitle
%%\tableofcontents
%%\newpage
\section{Introduction \label{sec:intro}}

In the search for physics beyond the Standard Model (SM), the top sector holds unique significance. As the top quark has the largest Yukawa coupling to the Higgs, it contributes one of the largest loop corrections to the Higgs mass, exacerbating the Higgs naturalness problem. To avoid a large degree of tuning, we therefore expect a top partner \cite{Kitano:2006gv,Asano:2010ut} that is not too much heavier than the top itself, and can be considerably lighter than most other new physics states. 

In models of softly-broken supersymmetry (SUSY), this expectation is reinforced by the connection between electroweak symmetry breaking and soft SUSY breaking. In any such model, one can write an expression that relates the mass of the Standard Model $Z$ boson to a linear combination of soft-breaking masses, together with the supersymmetric Higgsino mass parameter $\mu$. This implies either that the soft-breaking mass parameters are 
not too far above the electroweak scale, or that the underlying high energy theory enforces relations among parameters that lead to cancellations in
the effective low energy theory. However the latter option is itself strongly constrained by the renormalization group (RG) running of the soft-breaking SUSY parameters and SM parameters, which imply a complicated mapping from the high scale theory to the effective theory probed by experiments.
The largest RG effects are related to the largest couplings, and again the top sector has unique importance. This implies that one or both of the stop squarks, the scalar
superpartners of the top quark, are expected to be relatively light. 

In R-parity conserving SUSY the stop is not a good dark matter candidate, so we will neglect the possibility that the lightest stop is also
the lightest superpartner (LSP), and assume that the actual LSP is a weakly interacting particle such as a gaugino. 
It is quite possible, however, that the lightest stop ($\tilde{t}_1$) is the the next-to-lightest superpartner (NLSP). Because of $R$-parity and charge conservation,
stops are produced in pairs in hadronic collisons. Once produced, a stop will decay to the LSP plus SM particles, a decay that can be two body, three body, four body, or even more, depending upon the mass spectrum of the other superpartners whose off-shell couplings connect the stop to the LSP.

In the LHC era, null results from searches for extensions to the Standard Model have excluded new strongly interacting particles with masses that
in some cases exceed a TeV \cite{Chatrchyan:2011zy,Aad:2011ib}.
While the LHC experimental searches have been inclusive, the resulting mass limits vary according to the
production cross section and decay properties of the new particles. In particular squark mass limits derived from LHC experiments
often assume four flavors of degenerate squarks, with an additional two-fold degeneracy between the squark partners of left-handed and
right-handed quarks. These limits obviously do not apply to a single light stop. Both ATLAS and CMS have begun to constrain
models in which pair-produced gluinos decay via stop-top pairs \cite{Aad:2011ib,CMS-PAS-SUS-11-028}, but of course signals in this mode depend
on the gluinos being kinematically accessible.  Direct stop production has been constrained in the special case that both stops decay to a top and a neutralino, and the neutralino then decays to a gravitino and a $Z$; in this topology ATLAS excludes stops up to $240-330$~GeV (depending on the neutralino mass) using $2.05$~fb$^{-1}$ \cite{Aad:2012cz}.

In many models, including SUSY models (on which we focus our attentions), the top partner decays directly to a top and an
undetected weakly-interacting particle (\ie~$\tilde{t}\rightarrow t\,\chi$), leading to a final state with missing transverse energy ($\met$).
Our analysis will focus exclusively on this possibility, which is the most generic.
If there is a sufficiently light chargino then the decay $\tilde{t} \to b\,\chi^+$ becomes important, and we will
consider this important case in a sequel to this report.
Other special cases require more specialized consideration; for example light sleptons enhance stop decays
with multilepton final states. 
For stops lighter than the top, decays could proceed either through an off-shell top (a possibility we will consider in this work), an off-shell chargino, or through a flavor-changing decay $\tilde{t} \to c\,\chi$\cite{Djouadi:1996pi,Kraml:1996kz,Beenakker:1996de,CDF9834,Drees:2012dd}.  
The possibility of such very light stops is already constrained by Tevatron searches \cite{Aaltonen:2012tq,Abazov:2010xm}, but covering all of the
remaining parameter space at the LHC is challenging \cite{Carena:2008mj,He:2011tp,Bi:2011ha}.

For stop pairs decaying via  $\tilde{t}\rightarrow t\,\chi$ the current leading technique looks for excesses in $t\bar{t}+\slashed{E}_T$ with the top pair decaying into (semi-)leptonic final states \cite{Meade:2006dw,Han:2008gy}.  A recent study of the LHC reach suggests that semi-leptonic analysis could extend the bounds to $750-800$~GeV with 20~fb$^{-1}$, assuming that the lightest superpartner particle (LSP) is much less massive than the stop \cite{Bai:2012gs}. For heavier stops, the existing searches can be improved by using boosted top-tagging~\cite{Plehn:2010st,Plehn:2011tf,Plehn:2012pr}. However, as this requires a large splitting between the mass of the stop and the top+LSP pair, it is ineffective when the stop $p_T$ is below $\sim 200$~GeV.  In Refs.~\cite{Kats:2011it,Kats:2011qh}, it was estimated that, if updated to 1~fb$^{-1}$, LHC searches \cite{ATLAStop,Aad:2011wc} combined with previous searches at the Tevatron \cite{Abazov:2008kz,Aaltonen:2009sf,Aaltonen:2011rr,Aaltonen:2011na,Abazov:2012cz} could exclude direct stop pair production decaying to light gravitinos for stop masses up to 180~GeV. New results from stop searches with the full 2011 datasets are expected soon from ATLAS and CMS. Clearly, this is a search of considerable interest to the experimental and theoretical community; as a result, during the completion of this paper, we became aware of two additional theoretical groups working on the improving stop sensitivity at the LHC \cite{Han:2012fw,Kaplan:2012gd}.

Many theorists have considered the possibility that a light stop may be nearly degenerate in mass with the sum of the masses of its decay
products. Some have even proposed that ``degenerate'' stops are a natural result of well-motivated SUSY models.
For example in Ref.~\cite{Csaki:2012fh} an explicit model was presented with a nearly massless LSP and a lightest stop with mass 188 GeV.
The literature on degenerate stops has so far assumed that $m_{\tilde{t}} - m_{\chi} \simeq m_{t}$ implies that such particles are invisible
to $\met$-based LHC searches, even if the stops have rather large production rates. This implicit no-go theorem is especially strong for
stops decaying predominately via $\tilde{t}\rightarrow t\,\chi$, where the stop pair signal mimics conventional $t\bar{t}$ production.
Even away from the degenerate limit, semi-leptonic decay channels have the disadvantage that $\met$ from the LSPs 
has to compete with the $\met$ contributed by neutrinos from top decays.

In this report we dispel this conventional pessimism about  LHC detection of degenerate or nearly-degenerate stops, 
specifically for stops that are at least as heavy as the top quark.
We present search techniques that are sensitive to the pair production of top partners decaying into tops and invisible particles, even in the case of exactly degenerate mass spectra. We consider both the semi-leptonic final state (isolated muon or electron plus hadronic jets plus $\met$) and the fully hadronic final state (jets $+$ $\met$). For the semi-leptonic case we assume a conventional lepton trigger, while for the hadronic final state
we assume a four-jet trigger as already implemented by CMS and ATLAS \cite{ATLAS-CONF-2012-037}.

Our first major observation is that the $\met$ distribution for stop pair production differs significantly from that of $t\bar{t}$,
even in the case where the stops are degenerate. This follows from the fact that $\met$, despite its calorimeter-centric origins,
is a measurement of missing momentum, not missing energy, as well as the fact that stops and tops have a significant decay width.
The resulting intrinsic differences in $\met$ for stops and tops are then magnified by boosted kinematics, taking advantage of
the large phase space accessible to stop and top production at the LHC.

Our second major observation is that even rather small differences in $\met$ or $\met$-related spectra can be 
detected using a shape analysis. For $\met$-based observables we show that, in the kinematic regime where signal is enhanced, the distributions of the relevant backgrounds are well-described by simple analytic functions.
This background-fitting technique is motivated by the 
CMS Razor searches \cite{PhysRevD.85.012004,CMSrazor,CMS-PAS-EXO-11-030}, which in 2010 and 2011
successfully implemented one and two-dimensional shape analyses into inclusive SUSY searches and a third-generation
leptoquark search. The Razor searches are
based on the Razor kinematic variables $M_R$ and $R$, where $R$ is related to the $\met$ fraction of the event \cite{Rogan:2010kb}.
Rather than attempt to reproduce the 2D Razor fitting techniques, our analysis focuses on simpler 1D shape analyses.
The success of the Razor both validates the realism of our basic approach, and suggests that the
application of the CMS Razor to a degenerate stop search would result in equal or greater sensitivity than discussed here.

In Section~\ref{sec:method}, we describe in detail our search strategy, focusing on the missing transverse momentum distribution of stop events, as well as the fitting of background distributions. In Section~\ref{sec:hadresults}, we use the results of the {\tt MadGraph}, {\tt Pythia}, and 
modified {\tt PGS4} simulation tools to demonstrate the reach of our technique for hadronic stop searches in the next year of LHC running. Finally, in Section~\ref{sec:semilep}, we apply the shape analysis technique to a kinematic distribution related to $\met$ ($M_T^W$) in semi-leptonic top decays. In our conclusions, we present the expected exclusion limit using the combination of these two orthogonal searches; as we will show, using the expected luminosity from the LHC in 2012 (20~fb$^{-1}$) these shape analyses can potentially exclude stops up to $560$~GeV when the LSP is very light, and up to 360~GeV when the sum of the top and LSP masses are degenerate with the stop. This constitutes a significant  improvement over the current and projected bounds from the standard stop searches.

%%%%%%%%%%%%%%%%%%%%%%%%%%%%%%%%%%%%%%
\section{$\met$ and $\met$-related Methods} \label{sec:method}

As discussed above, the SUSY scenario we wish to consider is one in which the stop is considerably lighter than the other squarks (and gluino) and decays directly to a top and the LSP, 
which for concreteness we take to be a neutralino.  In particular, we will consider the simplified model \cite{Alves:2011wf}
of a single light stop squark ($\tilde{t}_1$, henceforth $\tilde{t}$) which decays to a top and a neutralino.  Since the stop is a colored scalar, its production is dominated by QCD processes and so is only very weakly sensitive to the details of the rest of the superpartner spectrum

\begin{figure}[t]
\includegraphics[width=0.4\columnwidth]{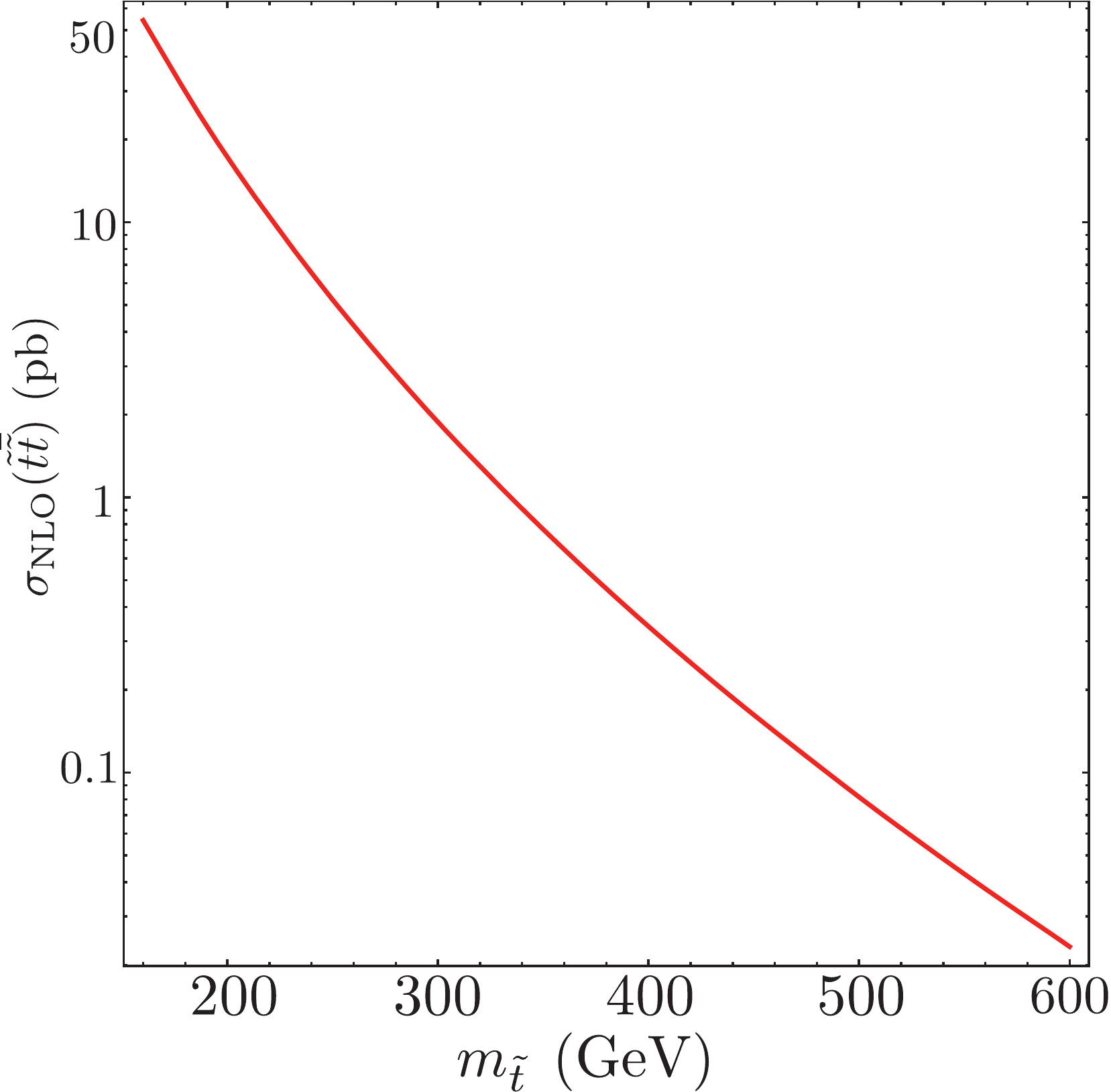}
\caption{Stop pair production at $\sqrt{s}= 8$~TeV, calculated at NLO using {\tt Prospino}~\cite{Beenakker:1996ed}. \label{fig:cross_section}}
\end{figure}

The rate and kinematics of the $\tilde{t}\rightarrow t\chi$ process are determined by two parameters: the stop mass, $m_{\tilde{t}}$, and the neutralino mass, $m_\chi$.  Once the stop decays, and the neutralino escapes the detector the only visible states in the signal events are the decay products of the tops. The only remaining indication of stop production is the missing transverse energy carried away by the LSPs.  In Figure~\ref{fig:cross_section} we show the production cross section at NLO for LHC with $\sqrt{s} = 8$~TeV. All other superpartner masses are set to 1~TeV, except for the neutralino. Clearly, the small production cross section for a single stop pair, combined with the lack of multiple observables to distinguish from background makes the search for stops challenging.  

However, the presence of intrinsic $\met$ is a handle that allows signal to be distinguished from backgrounds.  The existence of $\met$ in stop events will not only affect the number of events with large $\met$ but also the \emph{distribution} of these events. Furthermore, the background distribution of $\met$ can be well modeled using simple analytic functions, which can, in many cases be measured in high statistics control regions.  Using the shape of the $\met$ distribution provides a powerful tool to distinguish signal and background, as we outline below.

%%%%%%%%%%%%%%%%%
%%%%%%%%%%%%%%%%%

\subsection*{Background $\met$ shapes}

Since the signal contains the decay products of two top quarks, and intrinsic $\met$, the largest SM backgrounds will come from $t\bar{t}$, QCD multi-jet production and $W+$ jets.  Which of these processes dominates depends on the range of $\met$ and the mode of top decay.  In order to limit the source of non-LSP $\met$ (which would dilute the signal), we consider only fully hadronic top decay, for which our analysis applies a lepton veto; and semi-leptonic top decays, for which we require exactly one isolated lepton.

For the case of fully hadronic tops (the full analysis of which is described in Section~\ref{sec:hadresults}), there are two major sources of $\met$ in top background events.  The first is from detector mis-measurement of top events where both $W$'s decay hadronically. The second case -- which dominates at large $\met$ -- is due to one or both of the $W$'s decaying into a $\tau$ which in turn decays hadronically.  In this case the $\nu_\tau$ present in the top decay provides an intrinsic source of SM $\met$.  Other sources of intrinsic $\met$ in hadronic SM events arise include neutrinos from heavy flavor decays, and events where one or both of the $W$'s decay leptonically and all charged leptons in the decay are lost, either due to acceptances or detector effects.  For the QCD background the dominant source of $\met$ is mis-measurement of the jets.

For semi-leptonic top decays (detailed in Section~\ref{sec:semilep}), the sources of $\met$ are similar, although as we now require one charged lepton there will be more neutrinos  (either from leptonic decays of the $W$ or leptonic decays of a $\tau$). The $W+$ jets process is a relevant (but subdominant) background for the semileptonic analysis and here the $\met$ comes from the leptonic decay of the $W$.  The main QCD contribution is from jets faking leptons, but the rate for this is low. In the background events with a leptonically decaying $W$, the transverse mass of the lepton and the $\met$ should lie below the $W$ mass; there is however a tail above the $W$ mass generated by events with a leptonic $W$ and a hadronic $\tau$. As we will show in the next section, this arrangement of background $\met$ allows for a significant increase in signal over background by combining $\met$ with other kinematic information into a transverse mass variable $M_T^W$.

\subsection*{$\met$ and $M_T^W$ distributions in signal}
%$\met$ and $M_T^W$

By looking in the hadronic channel with a lepton veto, the separation between events with intrinsic $\met$ (signal), and those with other sources of $\met$ (background), can be maximized.  One might expect that the stop signal missing transverse energy would also be very small, especially when the masses of the LSP and stop are such that $\Delta \equiv m_{\tilde{t}} - (m_t+m_\chi) \approx 0$, making separation difficult. However, it is important to remember that the name `missing transverse {\it energy}' is a misnomer. It is not the transverse energy that is measured -- rather the detectors record transverse {\it momentum}. In the rest frame of the parent stop, the momentum of the LSP is
\begin{equation}
Q =\frac{\sqrt{[m_{\tilde{t}}^2-(m_t+m_\chi)^2][m_{\tilde{t}}^2-(m_t-m_\chi)^2]}}{2m_{\tilde{t}}}~. \label{eq:LSPq}
\end{equation}
For small splitting the missing momentum scales as $Q\approx \sqrt{2\mu \Delta}$ if $\Delta  \ll m_\chi$ and $Q\approx \sqrt{\Delta (2m_\chi+\Delta )}$ if 
$\Delta \sim m_\chi\ll m_t$ (here $\mu$ is the reduced mass of the neutralino-top system). In all but the last case the scale of the missing momentum is enhanced above that of the small mass splitting, proportional only to the square root of the small mass scale.  

Even in the limit where the stop is completely degenerate with the top-neutralino system ($\Delta  = 0$), the decay will proceed through the stop (or top, though this possibility was neglected in the Monte Carlo methods used in this paper) being off-shell by an amount comparable to the width $\Gamma$.  In this limit, where we assume the decay is still prompt, $\Delta $ should be replaced with $\Gamma$ in the above expressions.
Thus, for stops produced $\gtap 5$~GeV off shell and $m_\chi\gtap 50\ \gev$, we expect the LSPs to carry $\sim 20$~GeV of momentum each, in the rest frame of the top.

The intrinsic $\met$ of the event is obtained from the vector sum of the LSP transeverse momenta in the lab frame. Each stop is not generically at rest in the lab frame, and is boosted with respect to the center-of-mass frame of the partonic collision. The presence of ISR activity also provides a transverse boost, and causes the tops and neutralinos resulting from the stop decays to not be back-to-back, increasing the $\met$. Taking all these effects into account, we expect a harder distribution of $\met$ in stop pair events than in $t\bar{t}$ events, even for degenerate stops . This is confirmed by explicit simulation, as we will show. Note that our detailed simulations with {\tt MadGraph} and {\tt Pythia} use
the matrix element for stop pair production plus an extra jet to more accurately model the effect on $\met$ of the stop pair recoiling
against an extra energetic jet.

For the case of semi-leptonic top decays, the background, as outlined above, also contains irreducible sources of $\met$.  However, in these cases there is a $\met$-related variable that distinguishes signal and background: the transverse mass $M_T^W$ defined below.  Though the visible decay products are identical in signal and background, we can try to distinguish the two by considering the difference between the invisible components. For signal, the $\met$ consists of two LSPs and a neutrino, while for background, it comes predominantly from a single neutrino, which partners with the visible lepton to form a $W$ boson. If we assume that all events come from SM $t\bar{t}$ events, and thus that the neutrino $p_T$ is equal to the observed $\met$, then we can attempt to reconstruct the $z$-component (up to a two-fold ambiguity) of the neutrino momentum, using the $W$-mass as a constraint:
\begin{equation}
p^\nu_z = \frac{p^\ell_z (M_W^2+2 \vec{p}_T^{~\ell} \cdot \vec{\slashed{E}}_T)\pm E_\ell \sqrt{(M_W^2+2\vec{p}_T^{~\ell} \cdot \vec{\slashed{E}}_T)^2-4(p_T^\ell)^2\slashed{E}_T^2}}{2p_T^\ell}. \label{eq:pnuz}
\end{equation}
Clearly, if the missing energy is either inaccurately measured or not due to a $W$-induced neutrino, then this reconstruction will fail. One indication of such a failure would be if the quantity in the square root can be negative. Defining 
\begin{equation}
(M_T^W)^2 \equiv 2(p_T^\ell \slashed{E}_T - \vec{p}_T^{~\ell} \cdot \vec{\slashed{E}}_T), \label{eq:mtw}
\end{equation}
we can improve the signal over background ratio by restricting ourselves to the region $M_T^W > M_W$. This improvement arises because only mis-measurement and hadronic taus can drive $M_T^W$ into this regime for background, while for signal, the vector sum of two neutralinos and the neutrino can easily result in $\met$ that satisfies this constraint, even without mis-measurement. 

\subsection*{Shape analyses}

Experimental analyses, particularly at hadron colliders, have tended to shy away from modeling the shape of MET
distributions. In final states dominated by jets, there is the complicated phenomenon of jet mis-measurement, or
more generally the nonlinear response of the calorimetry used for the standard calorimeter-based 
reconstruction of MET. However, the ATLAS and CMS experiments have already demonstrated the ability
to understand MET distributions in a variety of complex final states, and to simulate MET including the
contributions to MET from imperfect detector response and reconstruction \cite{Chatrchyan:1361632,Aad:2012re}.
Already in the 2010 LHC run, the Razor analysis at CMS demonstrated the usefulness of modeling MET-based
observables for inclusive SUSY searches \cite{CMS-PAS-SUS-11-008}, and a similar approach was applied in the 2011 run
to a Razor search for relatively light third-generation leptoquarks \cite{CMS-PAS-EXO-11-030}. The latter is especially relevant to the
search for light stops, since it involved $b$-tagging and was optimized for lighter particles producing 
weaker MET signals. These successful shape analyses in jet-dominated final states in LHC data validate
that the basic approach pursued in this report can, with suitable modifications, be mapped into
successful searches.

For semi-leptonic final states there is an even stronger track record of successful modeling of MET-based
observables. In particular, the spectacularly precise determinations of the $W$ boson mass by the CDF
and D0 experiments were based on modeling of the $M_T^W$ distribution in large lepton-triggered
data sets \cite{TevatronElectroweakWorkingGroup:2012gb}. For stop searches we require much less precision in the determination of the shape,
and we are interested exclusively in modeling the tail above the Jacobian peak, rather than the peak itself.

For our study we have relied on simulated samples, with events generated by {\tt MadGraph5}~\cite{Alwall:2011vn}, showered and hadronized
by {\tt Pythia6}~\cite{Sjostrand:ys}, and physics objects reconstructed using {\tt PGS}~\cite{PGS}. The use of {\tt MadGraph} allows us to
simulate both SUSY signals and the $t\bar{t}$ background with extra partonic jets included in the matrix element.
This adds essential realism both in that initial state radiation (ISR) effects are important when simulating 
degenerate stops, and because our baseline selection relies on counting jets. {\tt PGS} has been shown to
give reasonably accurate results for MET and other basic observables for the case of SUSY signals \cite{Hubisz:2008gg,LL}
and, by extension, $t\bar{t}$ as long as one does not probe too far out in the tails of distributions.

Accurate simulation of QCD multijet backgrounds and the MET associated with them
is a more serious challenge, both because of the difficulty
of generating samples with sufficient Monte Carlo statistics, and trusting features of such samples in a toy
detector simulation after making very hard cuts. For our analysis we generated the equivalent of approximately 2~fb$^{-1}$ of QCD multijets.
We used the loosest of several different baseline selections (all requiring $b$-jets tagged to varying degrees of strictness) that seemed to give roughly comparable
sensitivity, with the idea that this makes our background modeling more reliable. 
While we have some confidence that our results agree at least qualitatively with distributions
obtained from LHC data, our simulated background samples should be considered as placeholders
for data control samples in a real LHC analysis. 

For this study we simulated the three largest backgrounds: QCD multijets, $t\bar{t}$+jets, and $W$+jets, but
neglect the smaller contributions from $Z/\gamma^*$+jets, dibosons, single top, and $t\bar{t}$+$Z$.

%%%%%%%%%%%%%%%%%
\begin{figure}[t]
\includegraphics[width=0.5\columnwidth]{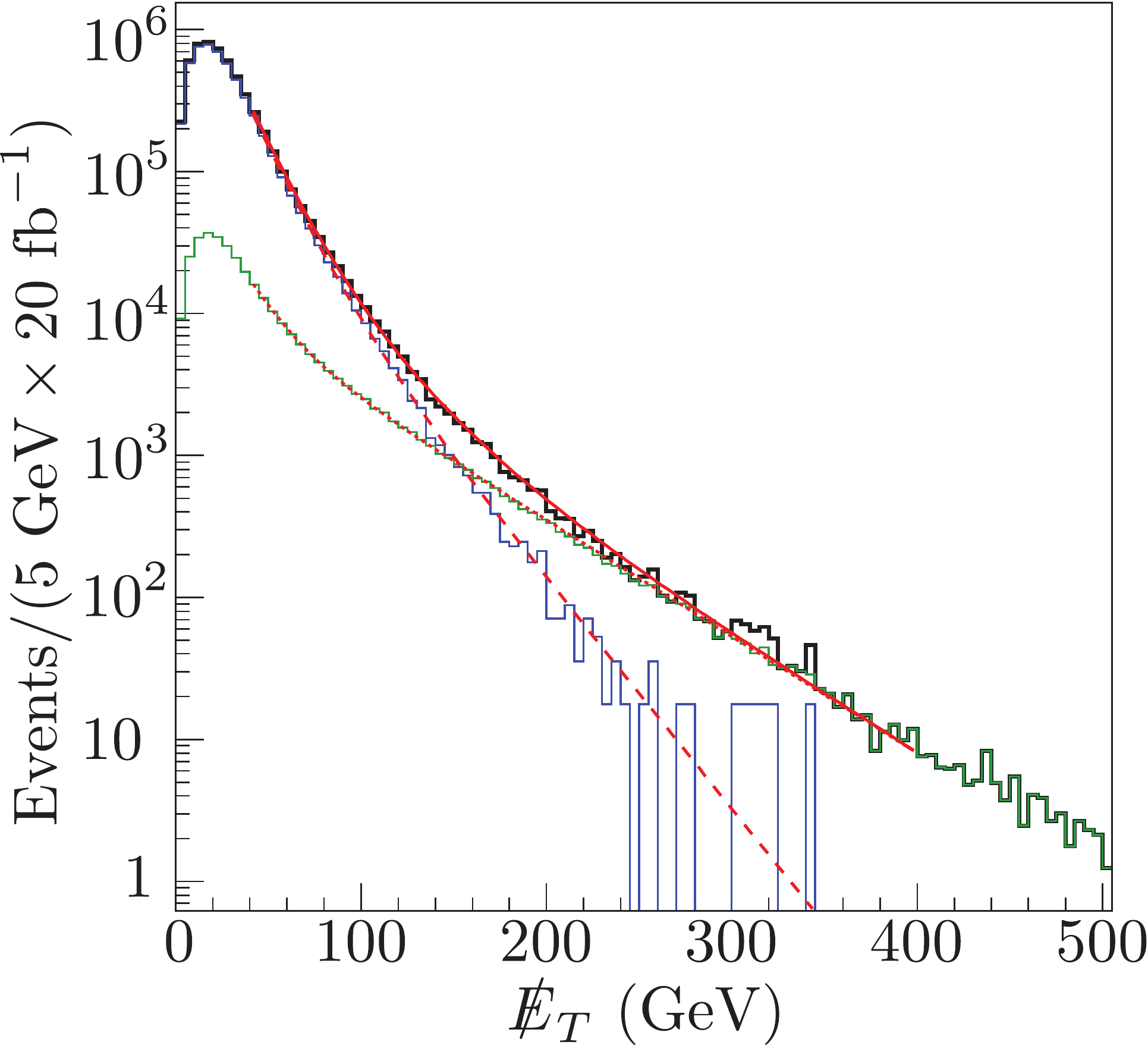}\includegraphics[width=0.5\columnwidth]{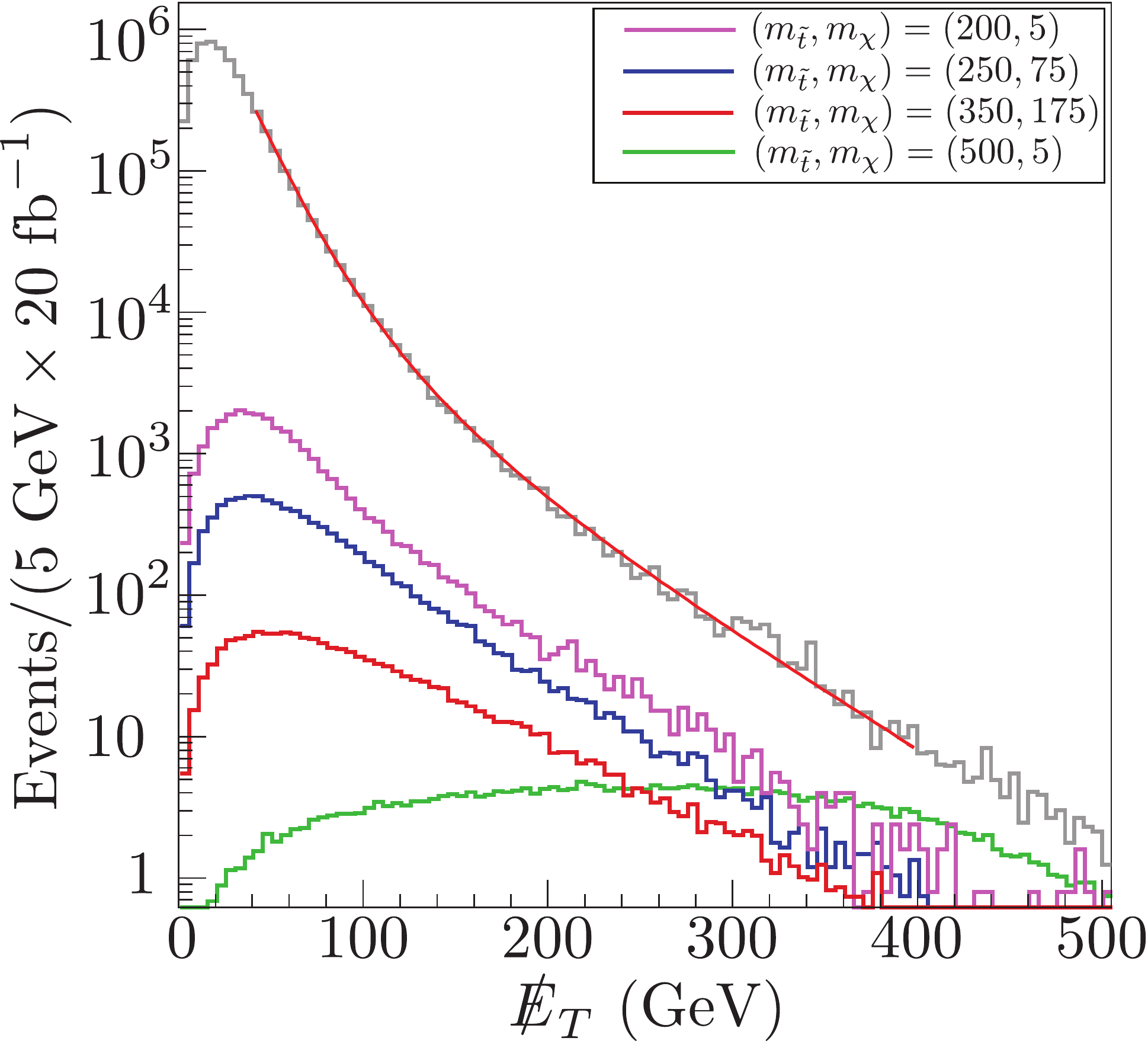}
\caption{Left: Differential distribution of events for 20~fb$^{-1}$ with respect to $\met$ of QCD (blue) and $t\bar{t}$ (green), and the total background (black) passing the hadronic trigger. The analytic fits to Eq.~\eqref{eq:METfit} using the parameters in Table~\ref{tab:bestfit} are shown in red for QCD (dashed), $t\bar{t}$ (dotted) and their sum (solid). Right: Differential distribution of events corresponding to 20~fb$^{-1}$ with respect to $\met$ for signal $\tilde{t}\bar{\tilde{t}}\to t\bar{t}\chi\chi$ passing the hadronic trigger for a range of stop and LSP masses $(m_{\tilde{t}},m_\chi)$. \label{fig:top_background}}
\end{figure}
%%%%%%%%%%%%%%%%%

Figure~\ref{fig:top_background} (left) shows the background MET distributions that we obtain after our hadronic baseline selection (detailed in Section~\ref{sec:hadresults}). QCD multijets dominates for MET values below about 150 GeV, while $t\bar{t}$ dominates above.
$W$+jets and other backgrounds were found to have a negligible effect on the MET shape above about
40 GeV. Above 40 GeV, both the QCD and $t\bar{t}$ backgrounds have MET distributions with a simple shape.
Both shapes are well-described by the sum of two exponentials, a feature reminiscent of the kinematic
shapes in the Razor analyses. The results of a fit (from RooFit \cite{Verkerke:2003ir}) in the MET range between 40 and 400
GeV are shown in red in the figure. The MET distributions of the hadronic signals from light stops also have
simple shapes, as illustrated in Figure 2 (right). As expected, while the signals suffer from lower cross
sections compared to background, for MET exceeding  $\sim 100$ GeV they start to emerge as significant
distortions of the MET shape. For degenerate stops the signal MET shapes have an exponential drop-off
that is similar -- but not identical -- to that of $t\bar{t}$. 

One could employ a more traditional ``cut and count" approach to the light stop analysis, but it is clear
from Figure~\ref{fig:top_background} that such an analysis would be complicated by the variety of different signal shapes
and signal MET regions of interest. However, it can serve as a useful cross-check, and so (as we will show), we have performed a simple cut-and-count analysis for comparison
to our shape analysis. An intermediate approach is to replace our analytic fits to the background shapes
with a coarsely-binned analysis of MET yields; however given the simplicity of the background shapes
it is not surprising that such a ``poor-man's" shape analysis has less sensitivity when compared to the full shape analysis.

\begin{figure}[t]
\includegraphics[width=0.5\columnwidth]{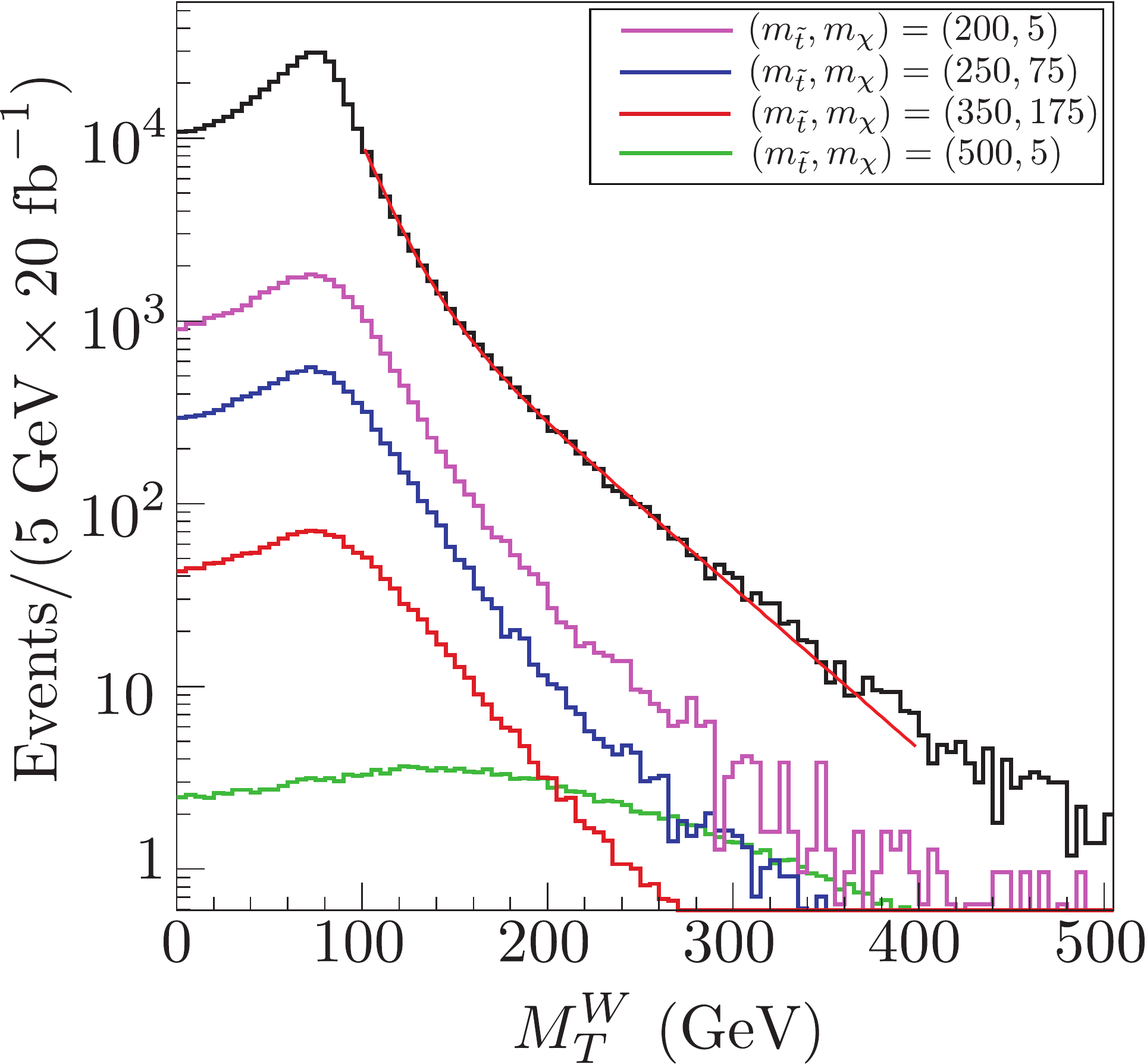}
\caption{Differential distribution of $t\bar{t}$ events with respect to $M_T^W$ (black). The analytic fit (Eq.~\eqref{eq:METfit} using the parameters of Table~\ref{tab:bestfit_lep}) is shown in red. Also shown are the differential distributions of stop signal events with respect to $M_T^W$ for a range of stop and LSP masses. The semi-leptonic event selection is described in Section~\ref{sec:semilep}.  \label{fig:mtw}}
\end{figure}

For the semileptonic analysis, the variable of interest is $M_T^W$ rather than $\met$; specifically, we are interested in the shape of $M_T^W$ above the mass of the $W$, where background is reduced. Using a lepton trigger followed by a tight $b$-tag, can significantly reduce $W+$jets background in this range, leaving only $t\bar{t}$ as the dominant background (the full baseline selection is described in Section~\ref{sec:semilep}). Using the same event generation as in the hadronic case, we show in Figure~\ref{fig:mtw} the distribution of $t\bar{t}$ background with respect to $M_T^W$. Above $M_W$, this distribution can, like $\met$, be fit with a pair of exponentials, greatly simplifying the shape analysis. Signal distributions for a representative sample of stop and LSP masses are also shown; as in the fully hadronic case, the shapes are sufficiently different to allow discrimination.

%%%%%%%%%%%%%%%%
%%%%%%%%%%    
%%%%%%%%%%%%%%%%%%%%%%%%%%%%%%%%%%%%%%%%%
\section{LHC Search for Hadronic Stops \label{sec:hadresults}}

In order to look for the effects of stops in the shape of $\met$ in hadronic events, we must first significantly reduce QCD background. We do this by applying a baseline selection based on an all-hadronic trigger, simplifying those developed by ATLAS and CMS for LHC running. We require at least two jets with $p_T > 80$~GeV and at least two additional jets with $p_T>50$~GeV, with a requirement of $|\eta|<3$ for all jets. Of the jets with $p_T$ above $50$~GeV, two must be tagged as $b$-jets; at least one must pass a ``tight'' $b$-tagging requirement, and the second must pass at least the ``loose'' requirement. Events that contain any electrons with $p_T>20$~GeV, $|\eta|<2.5$ or any muons with $p_T>20$~GeV, $|\eta|<2.1$ are vetoed (for our simulations, taus are treated as jets, thus forming a irreducible background that contains large $\met$).

To calculate the efficiencies with which tops, QCD, and signal stop events pass the trigger, we perform Monte Carlo simulation of the CMS detector; using {\tt MadGraph5}/{\tt MadEvent}  to generate $t\bar{t}$ backgrounds and $\tilde{t}\bar{\tilde{t}}$ signal events, matched to one additional jet. The $\tilde{t}\to t\chi$ branching ratio is set to 1, and top decay is handled by {\tt Pythia6}. The top mass is assumed to be 175~GeV. Detector simulation is done by {\tt PGS4}, modified to more closely match the reported CMS $b$-tagging efficiencies for both ``tight'' and ``loose'' thresholds, as found in Ref.~\cite{CMSbtag}. 
The top cross section at $\sqrt{s} = 8$~TeV was calculated to be 226.9~pb using MCFM at NLO, while the stop cross sections were determined using {\tt Prospino}~\cite{Beenakker:1996ed}, and are shown in Figure~\ref{fig:cross_section}. 

%%%%%%%%%%%%%%%%%
\begin{figure}[t]
\includegraphics[width=0.4\columnwidth]{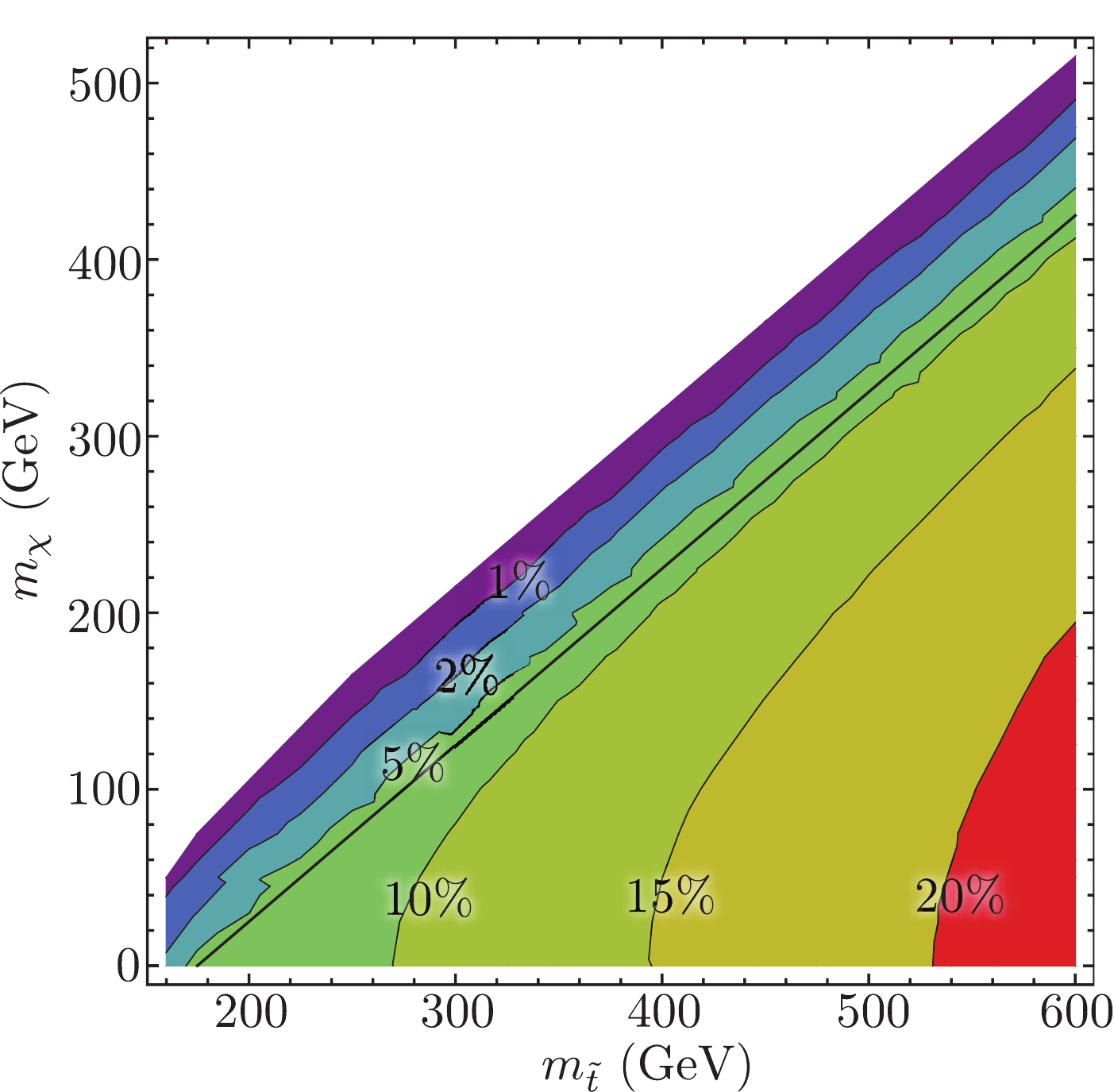}\includegraphics[width=0.4\columnwidth]{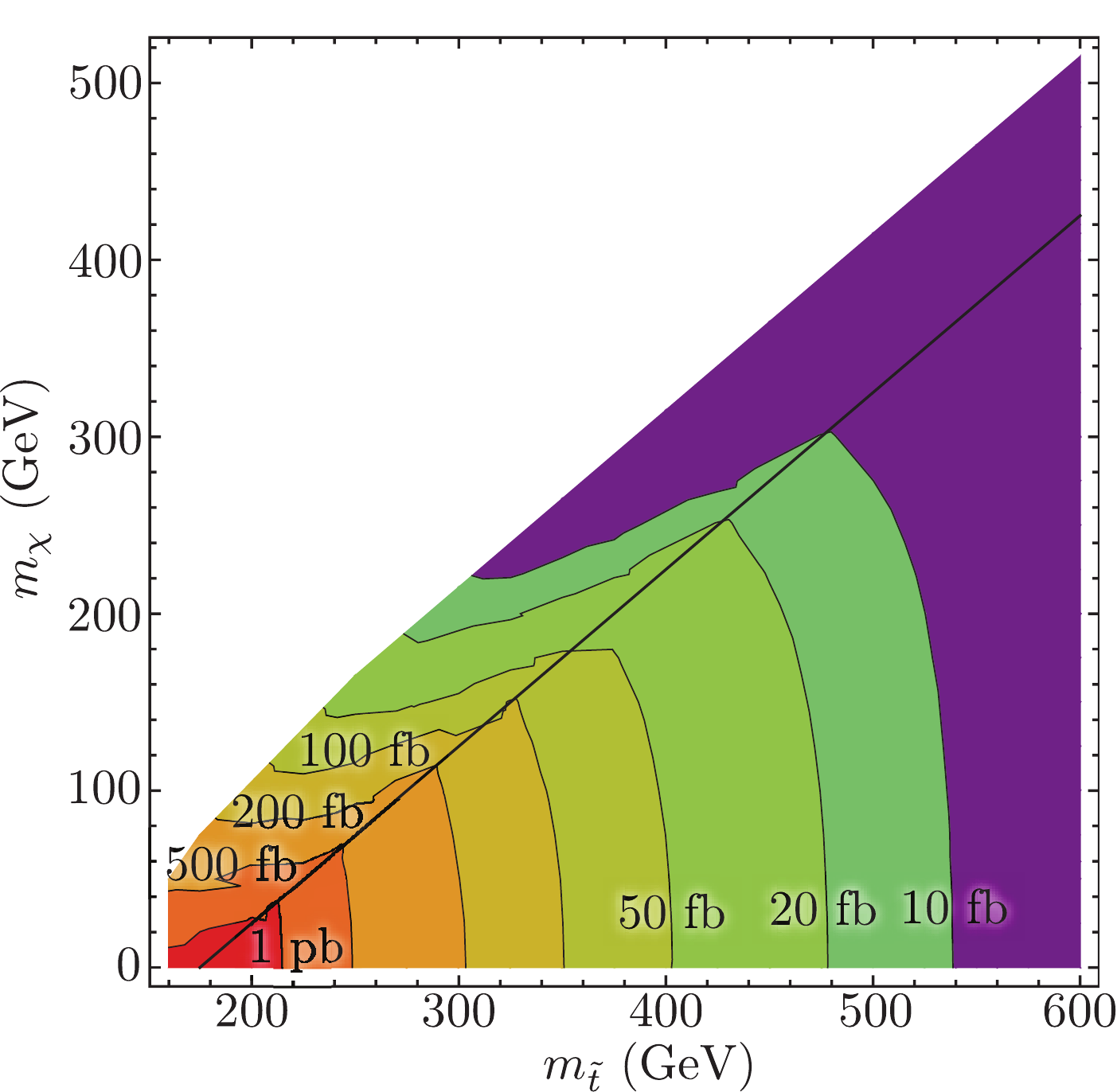}
\caption{Left: Signal trigger efficiency as a function of stop and LSP masses for hadronic event selection. Right: Signal cross section times trigger efficiencies as a function of stop and LSP masses. Like all such plots in this paper, the contours are extrapolated from a grid of Monte Carlo results with $5-25$~GeV spacing in $m_{\tilde{t}}$ and $m_\chi$. The degeneracy line ($m_{\tilde{t}}=m_t+m\chi$) is shown in black. \label{fig:hadeffs}}
\end{figure}
%%%%%%%%%%%%%%%%%

Using these simulations we find that the trigger has a $\sim  7\%$ pass efficiency for background tops, while the stop signal efficiency can vary from 2\%-20\% percent, depending on the mass splitting between the stops and the LSP (see Figure~\ref{fig:hadeffs}). Larger splittings lead to more energetic tops in the decay, and so result in more high-$p_T$ jets and a higher trigger efficiency. We generated four-jet QCD background events in {\tt MadGraph}, and allowed them to hadronize and shower through {\tt Pythia6}, which produced a higher multiplicity of jets. The QCD total cross section and differential rates at LHC7 were compared to ATLAS experimental results~\cite{Aad:2011tqa}, and scaled to LHC8 by taking the ratio of {\tt Alpgen}~\cite{Mangano:2002ea} partonic cross sections at LHC7 and LHC8. After application of our jet trigger selection and $b$-tag requirements, we find that $\sim 265$~pb of QCD background remains. However, only $17\%$ of these events have $\met$ above 40~GeV.

As stated previously, assuming perfect detectors and no contamination from events with leptons (and thus neutrinos), the top and QCD backgrounds should have zero $\met$. However, this is clearly not an assumption that survives contact with reality. Mismeasurement of jets, mis-tags of electrons and taus, and other experimental effects will all contribute non-zero $\met$ to the background. As we are limited to publicly available tools in our simulations, we cannot hope to exactly reproduce the $\met$ distribution in top events which will be observed by CMS and ATLAS. However, our {\tt PGS} simulation of the detector (using %$\met$ gaussian smearing of 0.2 and 
the CMS detector geometry) will be sufficient to demonstrate the general behavior.

In the left panel of Figure~\ref{fig:top_background}, we plot the $\met$ distribution for background events (in 5~GeV bins) passing our trigger selection criteria, using an initial set of 60 million QCD {\tt MadGraph}/{\tt Pythia}/{\tt PGS} events and 27 million top events. Two important features can be easily noticed. First, the $\met$ background peaks at $\sim 20$~GeV; this is at or below the intrinsic $\met$ value of stop events for all mass parameters of interest.  Second, past the peak, each background is exponentially falling.  We separately fit each background to a sum of two exponentials,
\begin{equation}
\frac{d\sigma}{d\met} = A e^{-\alpha\met}+Be^{-\beta\met}~. \label{eq:METfit}
\end{equation}

Due to limited statistics in the tail, and the complicated structure at low $\met$ , we only use this analytic fit over the range $40\le \met\le 400$ GeV. Other choices for the fitting function are possible (such as a Gaussian or Cruijff function, combined with exponentials), and may increase the range over which the background may be modeled. However, this simple choice suffices for our purposes. The corresponding distributions for signal are shown in the right-hand panel of Figure~\ref{fig:top_background} for a range of stop and LSP masses. For each signal point, we generate between 400,000 and one million matched stop pair events using {\tt MadGraph}/{\tt Pythia}/{\tt PGS}. However, we do not attempt an analytic fit for signal. Notice that, for signal, the total $\met$ peaks at a higher value than the parton-level $\met$ does. This is due to the addition of jet mis-measurement in addition to the LSP momenta, which serves to increase the average $\met$ observed. 

%%%%%%%%%%%%%%
\begin{table}[t]
   \centering   
   \begin{tabular}{|c|c|c|} % Column formatting, @{} suppresses leading/trailing space
\hline
	            &    QCD & $t\bar{t}$ \\
 \hline
     $\alpha$  &  $6.9\times10^{-2} \pm 1.56\times 10^{-3}$ &  $6.29\times10^{-2}\pm 1.63\times10^{-3}$\\
     $\beta$    &  $3.77\times10^{-2} \pm 1.26\times 10^{-3} $ &  $1.89\times 10^{-2}\pm 1.57\times 10^{-4}$\\
\hline      
     \end{tabular}
   \caption{Best fit parameters for QCD and $t\bar{t}$ $\met$ distributions, fit to (Eq.~\ref{eq:METfit}) for an integrated
   luminosity of 20 $fb^{-1}$.
   Note that these errors are correlated with each other and with the normalizations ($A,B$), which in turn depend on the amount of 
   integrated luminosity considered. See text for details.}
   \label{tab:bestfit}
\end{table}
%%%%%%%%%%%%%

Our analysis is based entirely on Monte Carlo (MC) simulated samples. As a result, in order to mimic the effects of statistical fluctuations one would expect to see in data, which will affect the precision of the fits, we carry out the fits outlined above on appropriately chosen samples of MC data.  For $t\bar{t}$ we can generate in MC the number of events expected after 20 fb$^{-1}$ of 8 TeV running and use this to extract the parameters.  For QCD we cannot hope to generate sufficient MC, so instead we carry out the fit on the 60 million QCD events that we have.  We then use this fit as an input to generate ``pseudo-data'' appropriate to 20 fb$^{-1}$, and refit to the pseudo-data.  This approach captures the uncertainty expected in the fit of real data. We show the best-fit slopes, and the associated errors, in Table~\ref{tab:bestfit}. However, note that there are sizable correlations between these fit parameters that need to be taken in to account when calculating the uncertainty on the fit.  

Although we are handicapped by having to rely on MC to determine the shape of the background distributions, the LHC collaborations do not suffer from this restriction, as they are in possession of copious amounts of data.  The QCD background to our signal contains two $b$-tagged jets, mostly a light quark or charm quark faking a $b$-quark, or from real $b$ production.  The complementary anti-b-tagged sample (4 jets above our cuts but with no $b$-tags), provides a clean sample of (predominantly) QCD events in which to measure the $\met$ distribution.  However, in order to extrapolate the $\met$ distribution from this sample to the signal region the $b$ mis-tag rate in QCD samples, as a function of jet $p_T$, must be well understood.  Through simulation we estimate that if this mis-tag rate is known to $\sim 20\%$ accuracy, as a function of $p_T$, then the effects on the determination of the parameters describing the QCD background are within our present uncertainties.  This is encouraging for a data-based analysis.  The $t\bar{t}$ background is harder to determine from data alone, but this issue is beyond the scope of our discussion.

\subsection*{Maximum likelihood method}

In order to estimate the potential for 20 fb$^{-1}$ of LHC8 data to exclude or observe the stop simplified model at a particular parameter point ($m_{\tilde{t}},\,m_\chi$), we must have some measure of the difference between signal and background $\met$ curves. The measure we employ is 
hypothesis testing with profiled likelihoods \cite{Cowan:2010js}. 
In this approach one calculates likelihoods assuming the observed data is the result of a particular hypothesis, maximizing the likelihoods over ``nuisance'' parameters, which in our case are the 8 parameters of the fits to the
background $\met$ shapes. We account for the known correlated uncertainties in the fit parameters by
introducing Gaussian penalty terms into the definition of the likelihoods.

Since the above procedure requires access to data, we instead ask the question of how well the experiments can \emph{expect} to do \emph{if} the data they observe is due to a particular model. There are two natural hypotheses that we can make for what the LHC may see: a) there is no light stop and the only production mechanisms are from the SM, or b) there is a light stop and the production cross section is as predicted in the MSSM.\footnote{There is clearly a continuum of possibilities: that there is a light stop and neutralino but the production cross section is different from what is predicted in the MSSM.  Carrying out a full scan in the stop production cross section is beyond the reach of this paper.} To calculate the likelihoods for these two hypotheses, we can take advantage of our background analytic function as well as the shape of the distribution of signal events, determined from MC, to generate pseudo-data which contains within it an equivalent amount of statistical fluctuation as 20 fb$^{-1}$ of actual data. We generate this pseudo-data using the central values of the best fit parameters found in Table~\ref{tab:bestfit_lep}).  We then attempt to fit this pseudo-data to both the SM only hypothesis and the SM+stop hypothesis.  

The log likelihood, including the constraint associated with the Gaussian uncertainties on the background fit parameters, $c_i$, is given by
\begin{equation}
\log L(c_i,\sigma) = \sum_{\mathrm{bins}} -\nu(c_i,\sigma) + n + n \log\left(\frac{\nu(c_i,\sigma)}{n}\right)-\frac{1}{2}\sum_{pq}\left(c_p-\bar{c}_p\right)C^{-1}_{pq}\left(c_q-\bar{c}_q\right)~, 
\label{eq:loglikely}
\end{equation}
where $\nu$ is the predicted number of events in a bin, $n$ is the observed number of events in a bin for a particular set of pseudo-data, $\bar{c}_i$ is the central value of the $i^{\mathrm{th}}$ fit parameter, and $C_{ij}$ is the covariance matrix of those fit parameters.  The second summation term in Eq.~\ref{eq:loglikely} is a constraint in the maximization, coming from assuming the uncertainties in the parameters of the background fit are Gaussian in nature. We allow the eight parameters involved in the background fits (four normalizations and four slopes) to vary within their uncertainties and maximize the log likelihood over these parameters and the signal production cross-section, $\sigma$. That is, for the SM only hypothesis, we maximize $\log L$ over $\hat{c}_i$ and $\hat{\sigma}$, and for the SM+stop hypothesis, we fix $\tilde{\sigma}$ to the NLO expectation, $\sigma_*$, and maximize $\log L$ over $\tilde{c}_i$. Since the pseudo-data was generated under the SM only hypothesis, $\hat{\sigma}\sim 0$ in all cases. 

As our test statistic we use twice the difference between these two values, 
\be
2\Delta \log L = 2\log L(\hat{c}_i,\,\hat{\sigma}) - 2\log L(\tilde{c}_i,\,\sigma_*)~,
\ee
which for clarity we convert into a number of standard deviations $n_\sigma = \sqrt{2 \Delta { \log L}}$. 
This $n_\sigma$ measures the incompatibility of the SM+stop versus SM only profiled likelihoods.
We repeat this process 200 times to obtain the average sensitivity.

In addition to the profile likelihood method described above we also investigate the sensitivity along the ``degeneracy line"  ($m_{\tilde{t}}-m_\chi = m_t$) using the CLs method \cite{Junk:1999kv,Read:2002hq}.  We do so by generating $10^4$ pseudo experiments under both background only and signal+background hypotheses and then use these pseudo experiments to determine the expected exclusion of signal, for an observation consistent with background.  Since the CLs method requires a high statistics sample of pseudo experiments we did not calculate the bounds for stop masses below $\sim 230\ \gev$.  For the median expected exclusion we assume that the log likelihood ratio of the observed data falls at the median of the background-only distribution.  For the one sigma CLs band we assume the data falls above/below the background median value by one sigma, and similarly for the two sigma band.

\subsection*{Estimated Hadronic Stop Bounds}

%%%%%%%%%%%%%%%%%%%%%%%%%%
\begin{figure}[t]
\centering{
\includegraphics[width=0.45\columnwidth]{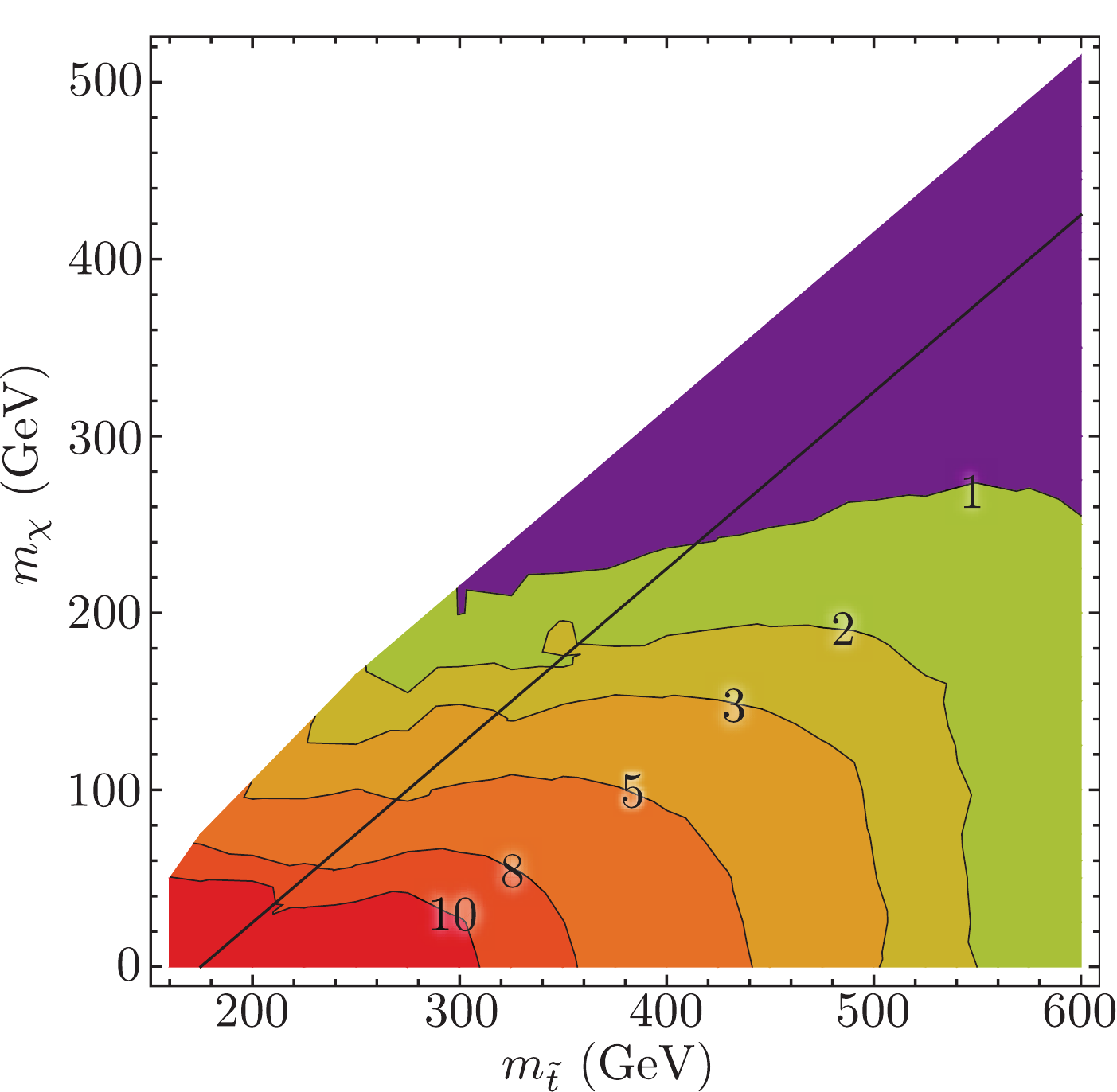}
\includegraphics[width=0.45\columnwidth]{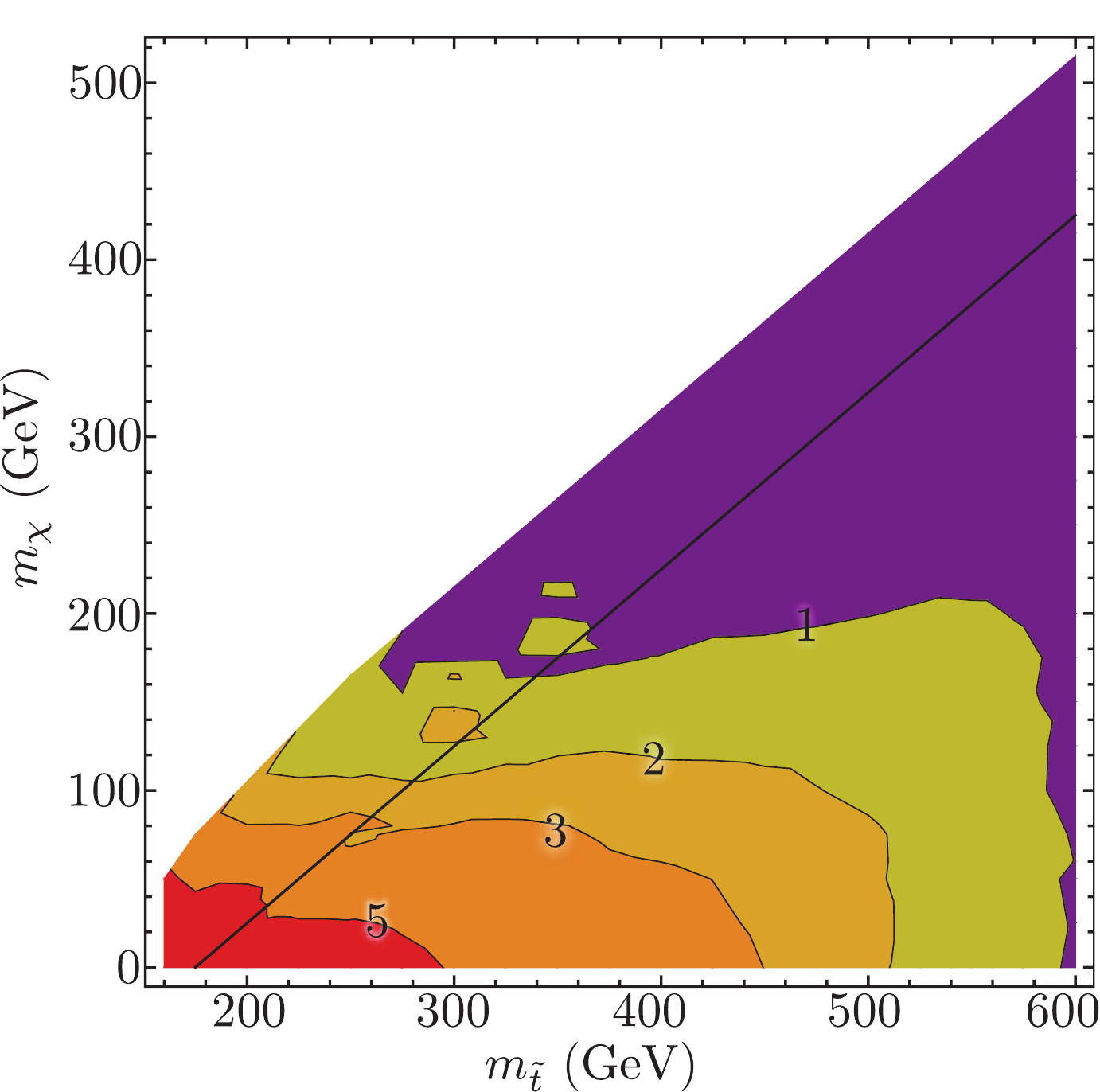}
}
\caption{Expected sensitivity, in standard deviations, for the hadronic $\met$ shape analysis as a function of the stop and LSP masses. The test statistic is computed from 200 pseudo-experiments of 20~fb$^{-1}$. In the left-hand plot the uncertainty on the background $\met$ shape are as shown in Table~\ref{tab:bestfit} and in the right-hand plot these errors have been inflated by a factor of 3.%Black points indicate mass parameters evaluated in Monte Carlo, contour lines are extrapolated from these point. See text for details. 
\label{fig:METexclusion}}
\end{figure}
%%%%%%%%%%%%%%%%%%%%%%%%

Using these statistical methods, in Figure~\ref{fig:METexclusion} we show the estimated significances extracted from our test statistic for light stop simplified models when the top decays hadronically.  %For comparison we have shown both the case where the fit uncertainties are taken into account and the unrealistic situation where the background shapes have zero error. 

We estimate that for simplified models in which the stop/neutralino mass splitting is large, the LHC experiments can set strong stop mass limits up to $\sim 550\ \gev$.  In the case of a very light neutralino the reach is determined simply by the production cross section of the stops, which drops rapidly with the mass (Figure~\ref{fig:cross_section}), although there is some softening of this behavior due to increased efficiency to pass the cuts as the stop mass is increased (Figure~\ref{fig:hadeffs}).

Most interestingly, even along the mass-degeneracy line of $m_{\tilde{t}}-m_\chi = m_t$, stops of mass as high as $\sim 350\ \gev$ could be excluded with 20~fb$^{-1}$ of 8 TeV data. In fact we find that the sensitivity reach extends above the degeneracy line into regions where the stops decay into off-shell tops.  

%%%%%%%%%%%%%%%%%%%
\begin{figure}[t]
\includegraphics[width=0.4\columnwidth]{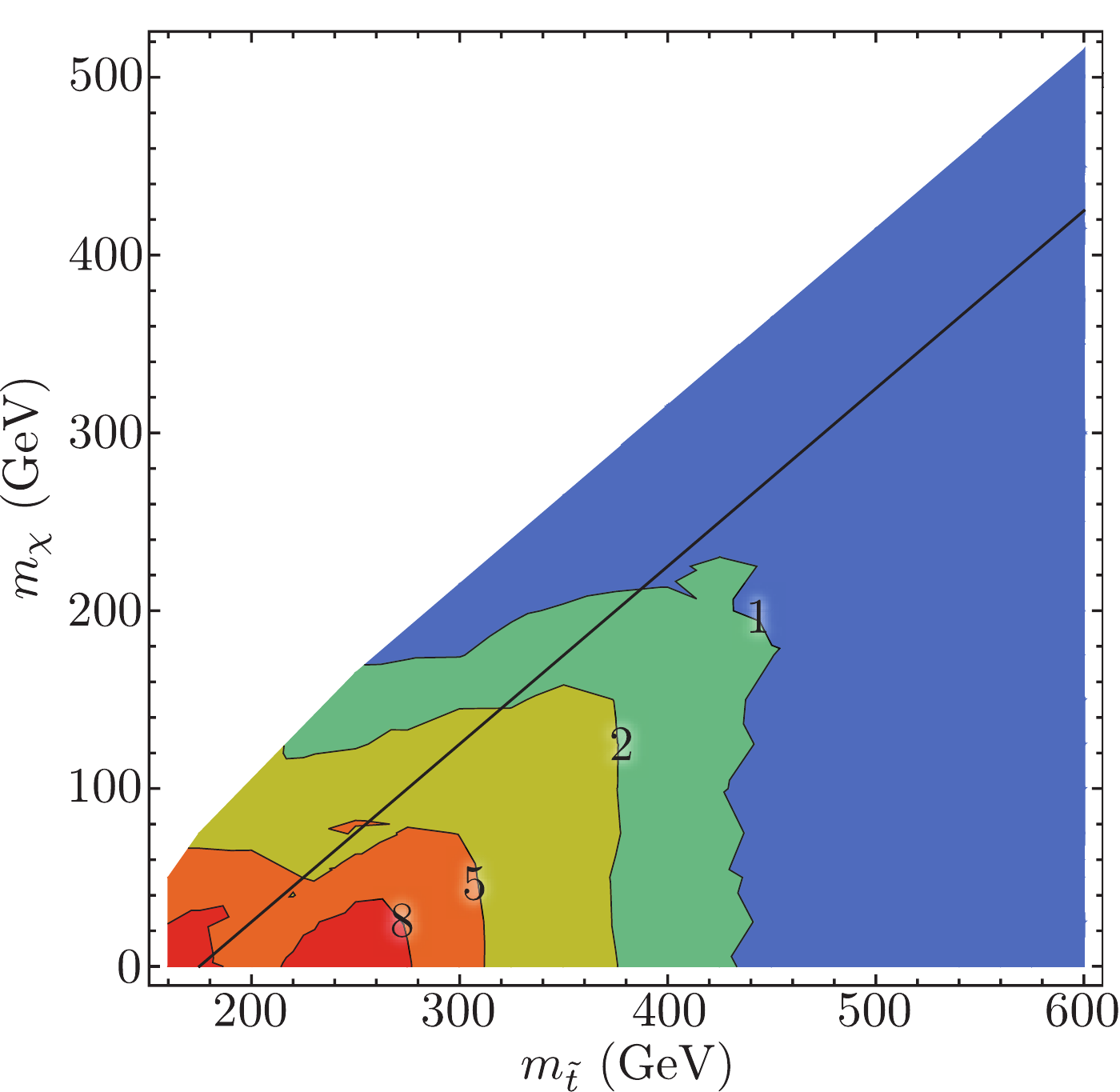}\includegraphics[width=0.4\columnwidth]{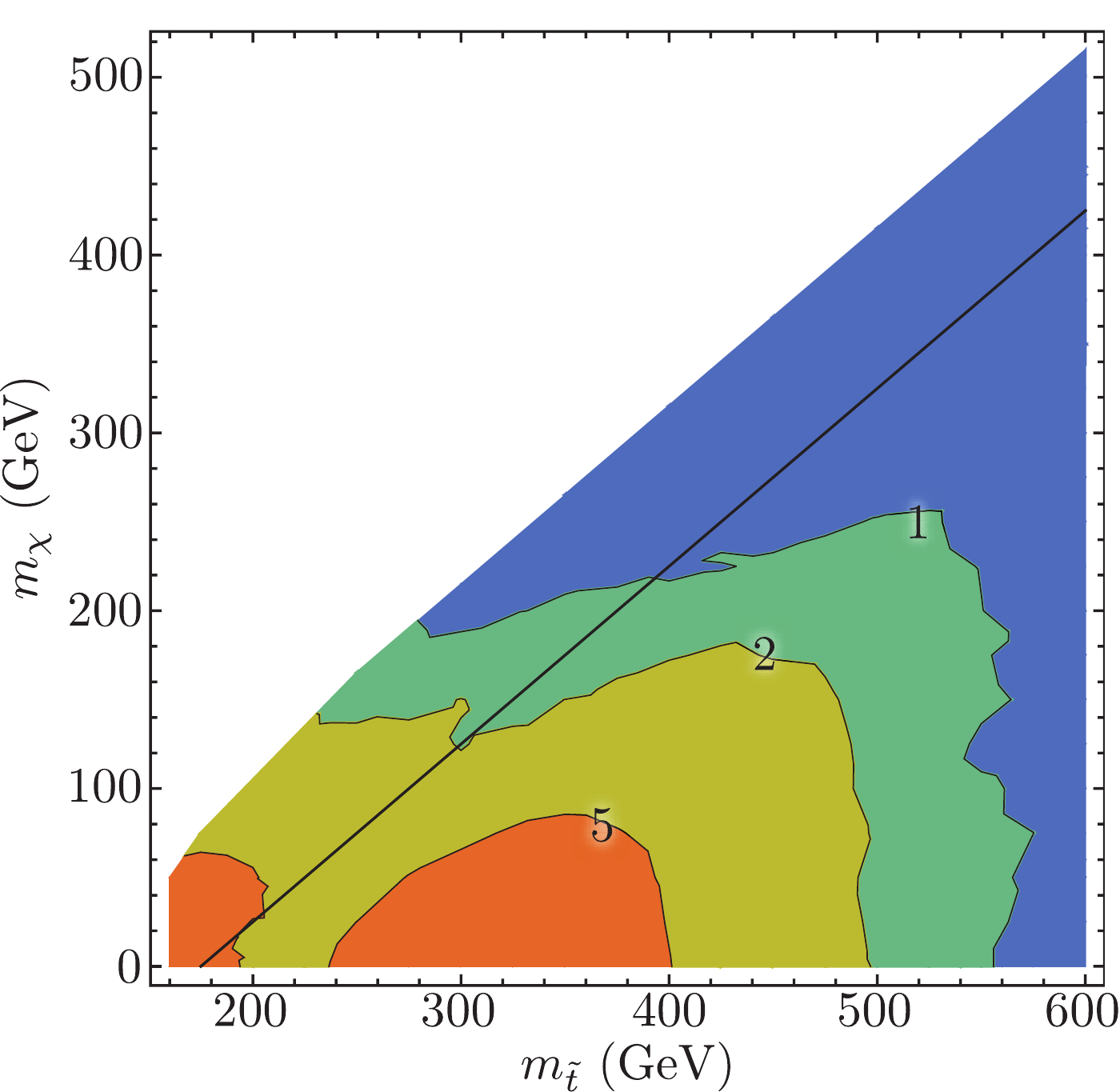}
\caption{Left: $S/\sqrt{O}$ for 100-200~GeV region of signal plus background $\met$ distribution. Right: $S/\sqrt{O}$ for $200-400$~GeV region.  These are computed with an unrealistic assumption of no systematics.\label{fig:METcc}}
\end{figure}
%%%%%%%%%%%%%%%%%%%%

As an additional cross-check, we perform a simple cut-and-count analysis of the signal parameter points, dividing the $\met$ range of 40-400~GeV into three regions: our ``background'' region of 40-100~GeV; and two signal regions; 100-200~GeV and 200-400~GeV. Iterating over 200 pseudo-experiments generating $\met$ distributions of signal plus background events, we assume that all events in the background region are ascribable to the QCD and top backgrounds. This sets our overall normalization, which we use to predict (using our analytic fit Eq.~\eqref{eq:METfit}) the number of the background events in our two signal regions. For each pseudo-experiment, we can then calculate the number of signal events $S$ in each signal region as the difference between the observed events $O$ and the predicted value $P$. In Figure~\ref{fig:METcc}, we plot the average value of $S/\sqrt{O}$ for both the low-$\met$ and high-$\met$ signal regions. 
Addition of a realistic systematic error to the predicted number of events will reduce the sensitivity of the cut and count method. For a stop mass of 250~GeV and LSP of 5~GeV one has, with zero systematics, 9$\sigma$ sensitivity, while a 5\% systematic reduces this to approximately 2$\sigma$. 

%%%%%%%%%%%%%%%%%%%
\begin{figure}[t] %  figure placement: here, top, bottom, or page
   \centering
   \includegraphics[width=0.6\columnwidth]{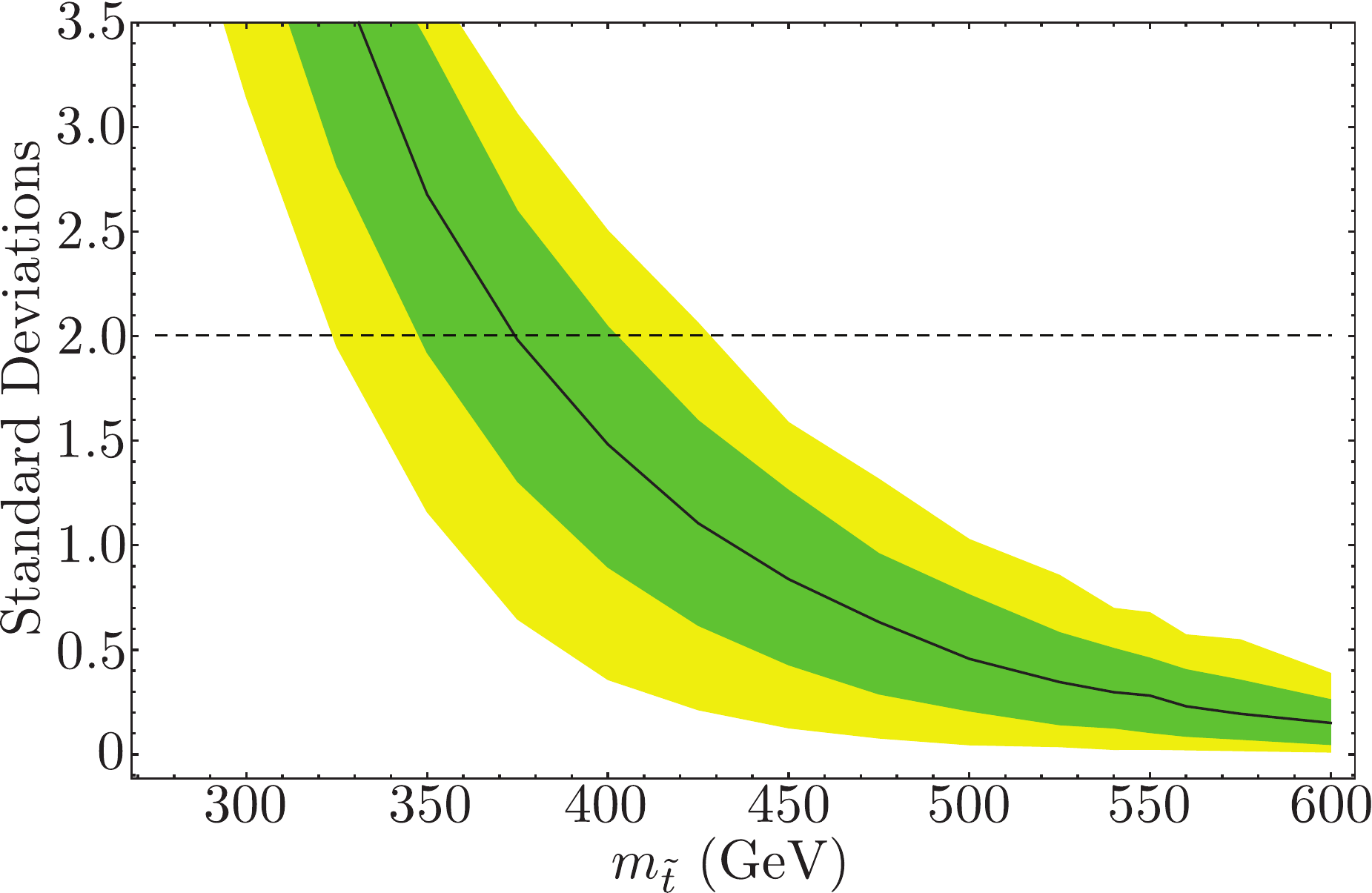} 
   \caption{The median expected exclusion, for background only pseudo experiments, on a stop-neutralino simplified model whose masses are related by $m_{\tilde{t}}-m_\chi = m_t$.}
   \label{fig:CLsplot}
\end{figure}
%%%%%%%%%%%%%%%%%%%

The shape analysis we are advocating allows for many of the backgrounds to be determined from control regions in data and thus removes many systematic uncertainties associated with theoretical predictions of background\footnote{There still exists a difficult-to-quantify systematic error associated with the choice of functional form the background distributions are fit to.  A discussion of these effects is beyond the scope of this paper.}.  There are systematic uncertainties associated with extrapolation from control regions to signal regions, such as the b-tagging rates discussed above, but we estimate that these are subdominant to the fit uncertainties.  We have, with the exception of QCD, fit to simulated data sets that are consistent with what one would expect after luminosity of 20 fb$^{-1}$ and find the errors as shown in Table~\ref{tab:bestfit}.  However, since our analysis is entirely MC based, and it is possible that the real control regions will contain limited statistics, we also investigate how the sensitivity is affected if the errors in our fit parameters are inflated.  In particular we consider the situation where the central values for the fit parameters are as shown in Table~\ref{tab:bestfit} but the errors are a factor of 3 or 5 times larger.  Assuming that the errors from extrapolation are then subdominant to the fit uncertainty, we keep the correlations between the fit parameters as we inflate the errors.  With an inflation by 3 the fractional errors in the fit parameters range from a few to 17\% and inflation by 5 has a largest error of 30\%, with the largest errors in the normalizations, as expected.  The effects of this inflation, for $3\times$, are shown in Figure~\ref{fig:METexclusion}.  An inflation by $5\times$ degrades the sensitivity as one moves towards degeneracy: along the degeneracy line the $2\sigma$ exclusion extends to $260\ \gev$.  The $2\sigma$ exclusion for case of light neutralino is not greatly altered from the bound for $3\times$ inflation.

Focusing on the degeneracy line ($m_{\tilde{t}}-m_\chi = m_t$), a region of particular interest and considerable challenge, we apply the CLs method as outlined above.  The median expected exclusion, as well as one and two sigma bands, on such a degenerate stop-neutralino pair is shown in Figure~\ref{fig:CLsplot}.   Using the CLs method, stop masses up to $375\ \gev$ can be excluded at $2\sigma$ when $m_{\tilde{t}}-m_\chi = m_t$.

%%%%%%%%%%%%%%%%%%%%%%%%%%%%%%%%%%%%%%%%%%%%
\section{Study of $M_T^W$ shapes \label{sec:semilep}}

We now turn from stops with fully hadronic decays of top to the semi-leptonic channel, discussed briefly in Section~\ref{sec:method}. In this case the dominant background is $t\bar{t}$. Given that semi-leptonic $t\bar{t}$ decays have an intrinsic source of missing transverse energy from the neutrinos coming from the $W$ decays, $\met$ offers poorer discrimination between signal and background, as compared to the hadronic case.
We therefore focus instead on the transverse mass variable $M_T^W$ defined in Equation \ref{eq:mtw}. This variable is related to
$\met$, but has the additional feature that SM background $\met$ from a single leptonic $W$ decay is mostly distributed below the
Jacobian peak near the $W$ mass. 

%%%%%%%%%%%%%%%%
\begin{figure}[t]
\includegraphics[width=0.4\columnwidth]{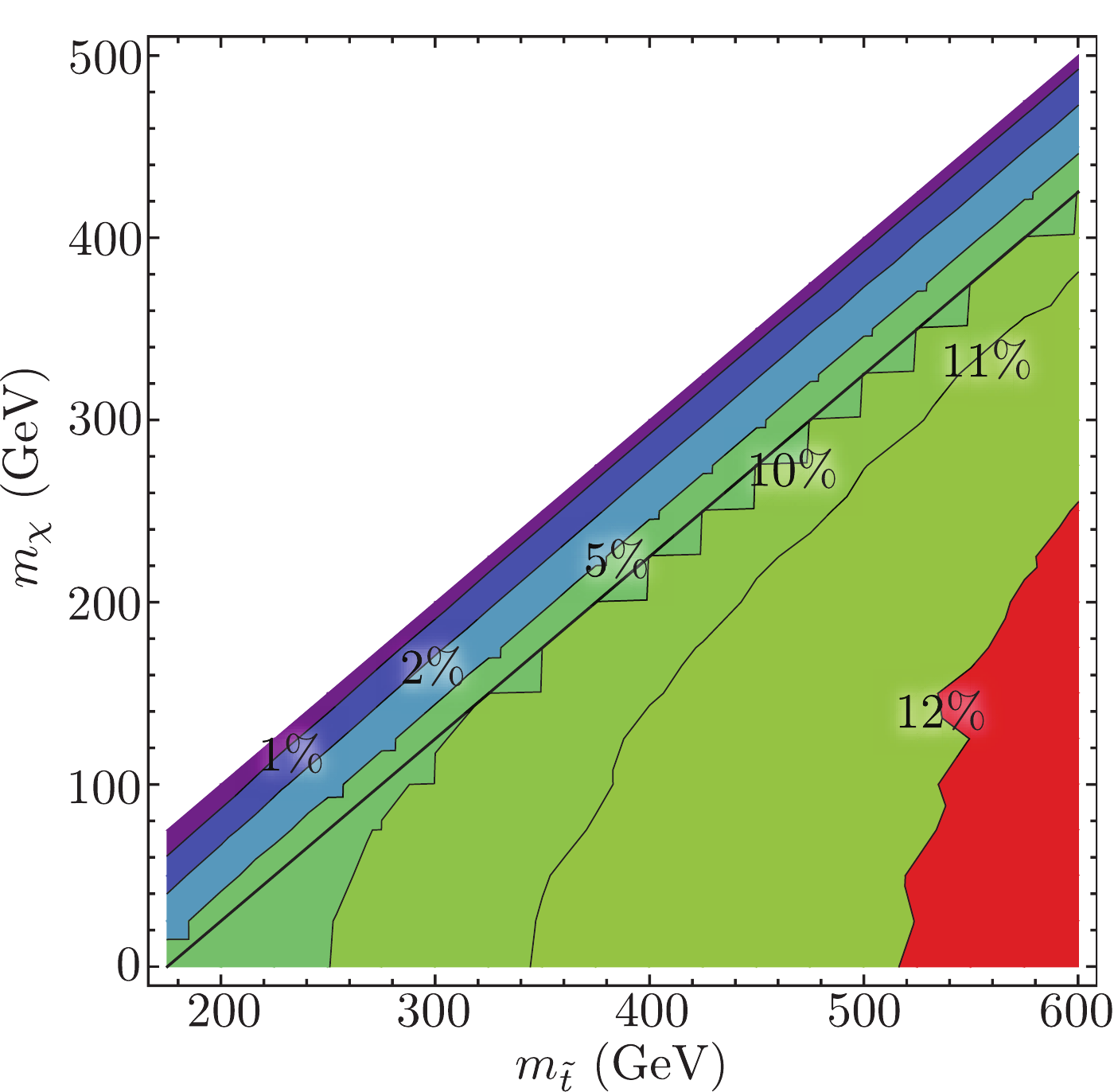}\includegraphics[width=0.4\columnwidth]{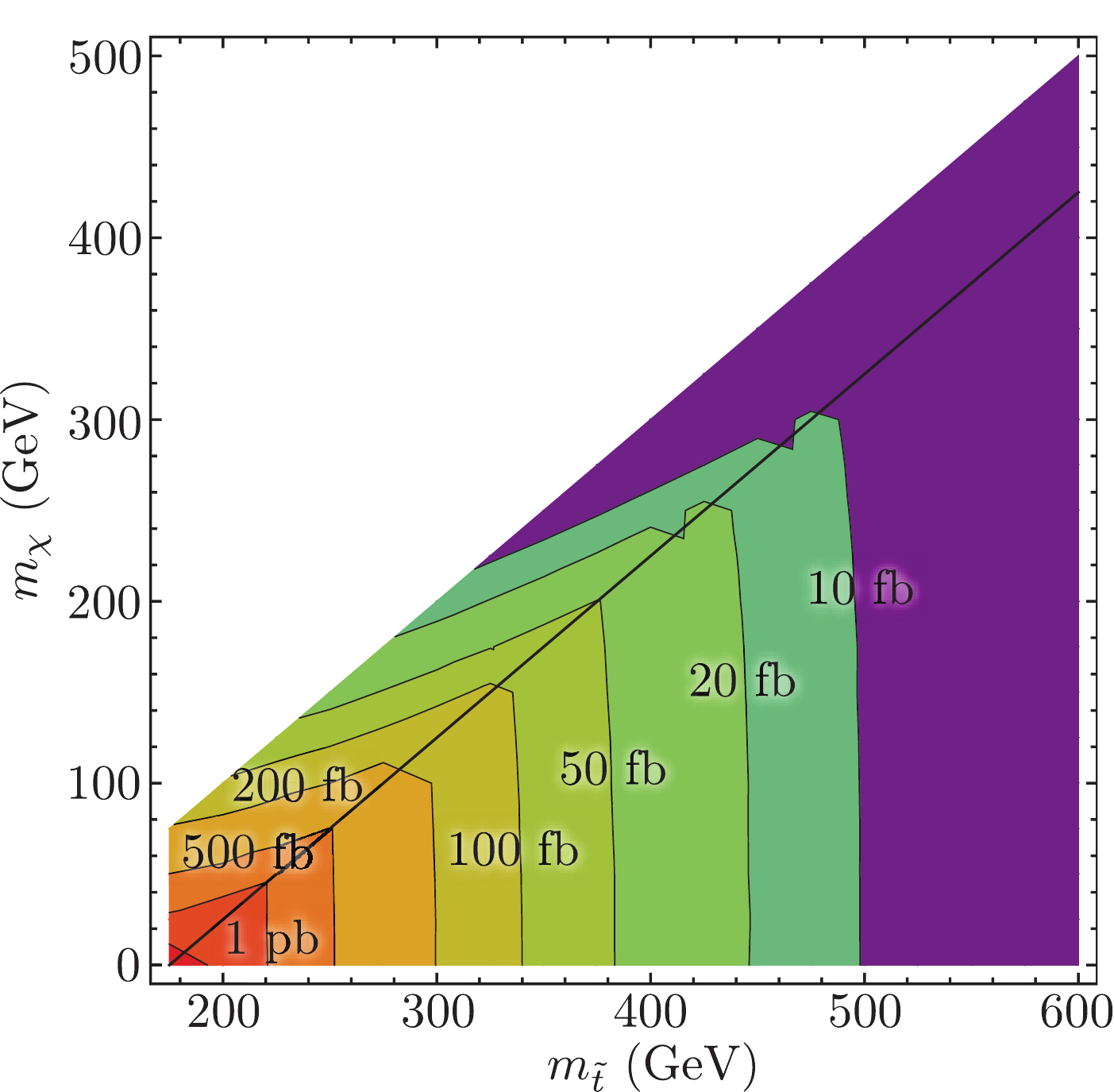}
\caption{Left: Semi-leptonic trigger efficiency for semi-leptonic events as a function of stop and LSP masses. Right: Cross section times efficiency for the semi-leptonic selection criteria as a function of stop and LSP masses. \label{fig:lep_eff}}
\end{figure}
%%%%%%%%%%%%%%%%%%%%%%

Our method follows the hadronic analysis closely. Again assuming stop pair production, each decaying to a top and an LSP, we now look for events where one top decays leptonically, while the other decays hadronically. We select events with exactly one isolated lepton with $p_T > 20$~GeV and $|\eta| <2.1(2.5)$ for muons (electrons), at least one tight $b$-tagged jet,  and requiring three or more jets with $p_T > 30$~GeV and
$|\eta| < 3$.
The primary background is reduced to $t\bar{t}$, with an acceptance efficiency of $\sim 15\%$ (including branching ratios). The efficiencies and cross section times efficiencies for the stop/LSP signal points are shown in Figure~\ref{fig:lep_eff}.

Focusing on $M_T^W$ above $M_W$ will improve the discrimination of stops from tops. Applying a shape analysis as was done in the hadronic $\met$ case will provide even greater advantages. The total SM background distribution for $M_T^W > 85$ GeV can again be well fit by the sum of two exponentials:
\begin{equation}
\frac{d\sigma}{d M_T^W} = A e^{-\alpha M_T^W}+Be^{-\beta M_T^W}. \label{eq:MTWfit}
\end{equation}
Repeating the search strategy performed in the hadronic analysis, we use {\tt RooFit} to find the best fit for the parameters in the $M_T^W$ range of $85-400$~GeV (see Table~\ref{tab:bestfit_lep}), weighting the top background to the equivalent of 20~fb$^{-1}$ of data. Again, the fit errors reported are highly correlated.

%%%%%%%%%%%%%%
\begin{table}[b]
   \centering   
   \begin{tabular}{|c|c|} % Column formatting, @{} suppresses leading/trailing space
\hline
	            &    fit to 20 fb$^{-1}$ total SM background\\
 \hline
     $\alpha$  &  $6.68\times10^{-2} \pm 6.88\times 10^{-4}$ \\
     $\beta$    &  $2.01\times10^{-2} \pm 3.04\times 10^{-3} $ \\
\hline      
     \end{tabular}
   \caption{Best fit slope parameters for background $M_T^W$ distribution, fit to (Eq.~\ref{eq:MTWfit}).  Note that the fit errors are correlated with each other and with the normalizations ($A,B$), which in turn depend on the amount of integrated luminosity considered.}
   \label{tab:bestfit_lep}
\end{table}
%%%%%%%%%%%%%

%%%%%%%%%%%%%%%%%%%%%%%%%%
\begin{figure}[t]
\centering{
\includegraphics[width=0.45\columnwidth]{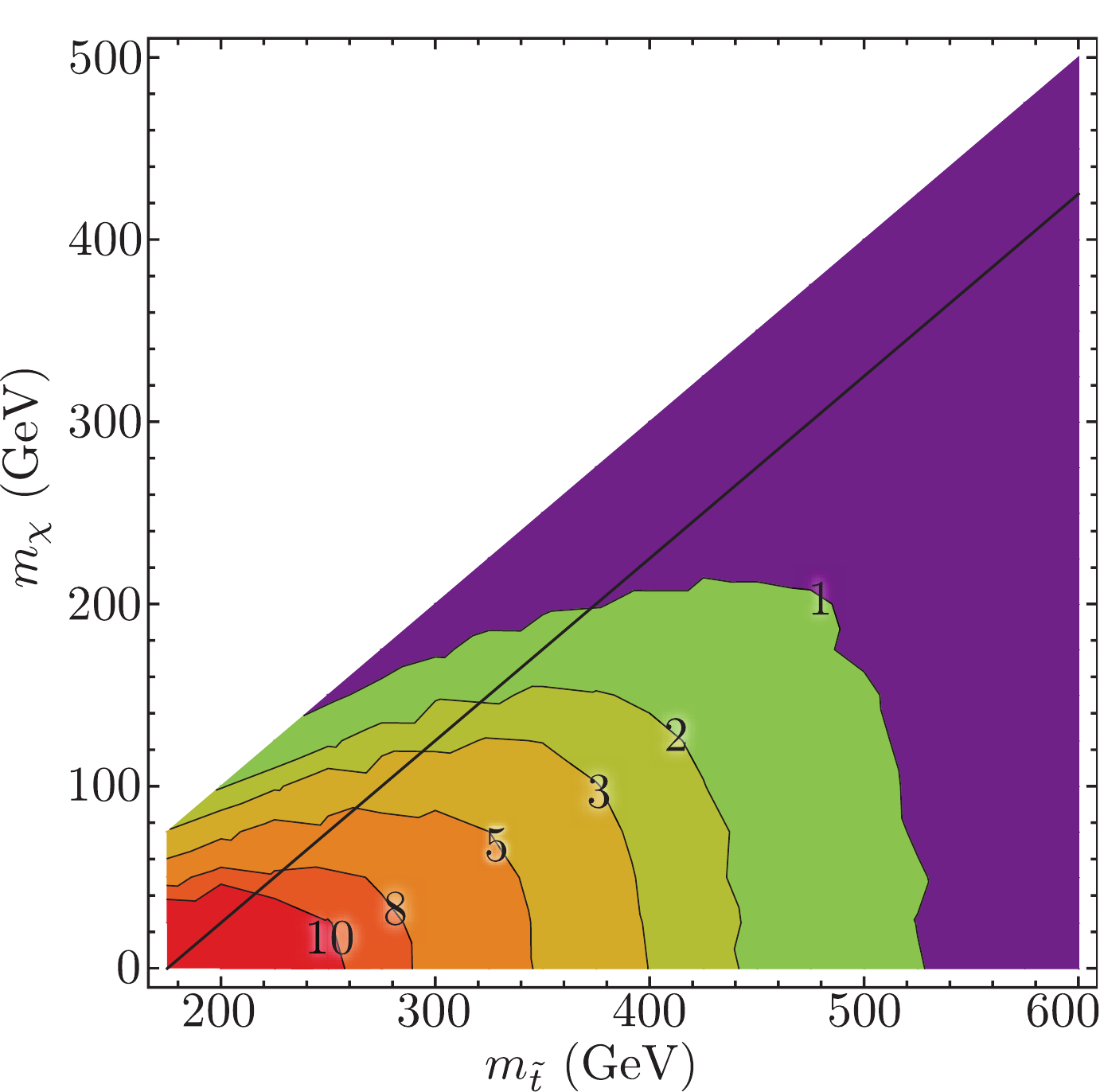}
\includegraphics[width=0.45\columnwidth]{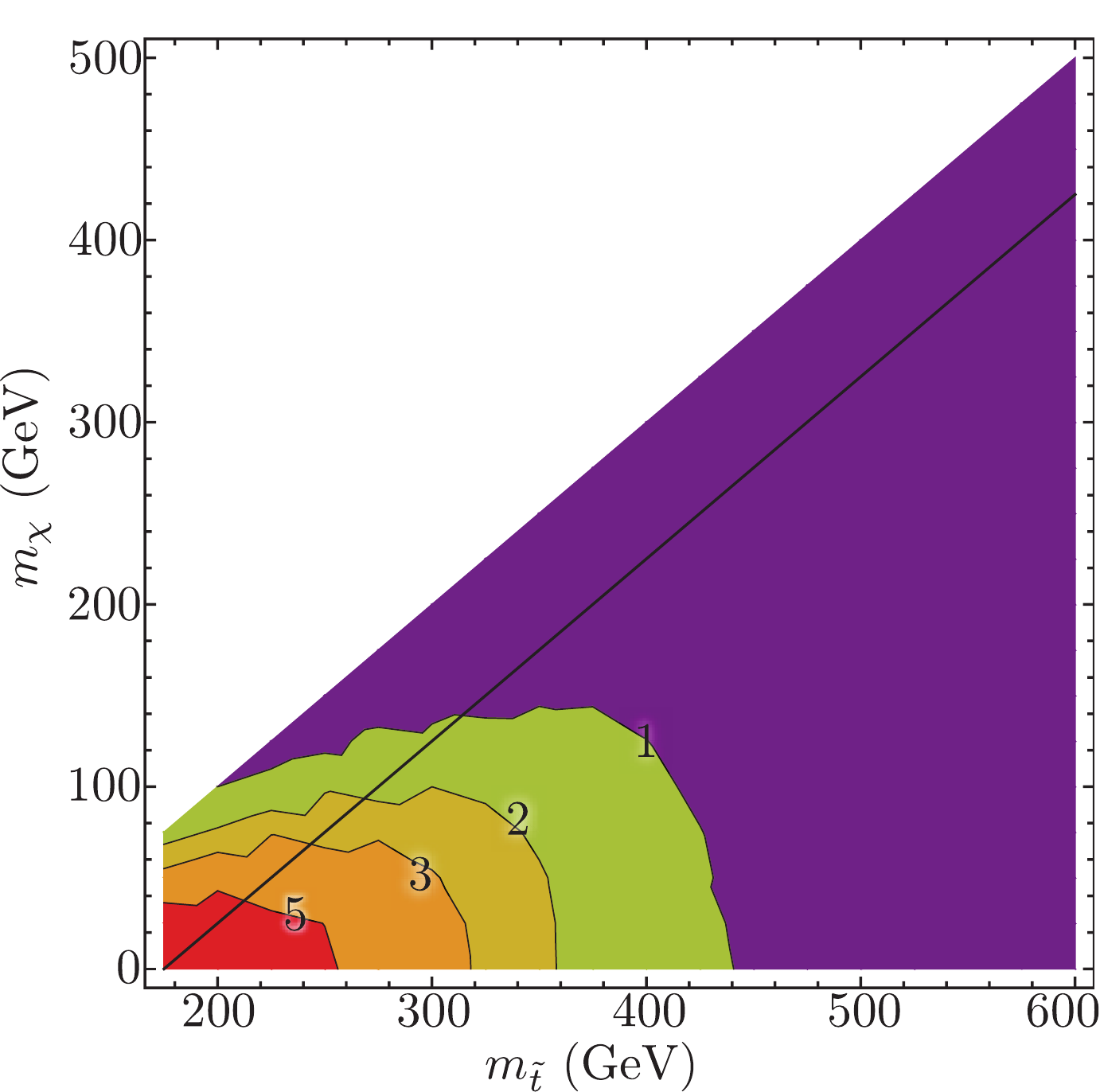}
}
\caption{Expected number of standard deviations that the supersymmetric stop signal can be excluded by using 200 pseudo-experiments of 20~fb$^{-1}$, applying the $M_T^W$ shape analysis.  In the left-hand plot the uncertainty on the background $\met$ shape are as shown in Table~\ref{tab:bestfit_lep} and in the right-hand plot these errors have been inflated by a factor of 3.
\label{fig:MTWexclusion}}
\end{figure}
%%%%%%%%%%%%%%%%%%%%%%%%

\begin{figure}[t]
\includegraphics[width=0.4\columnwidth]{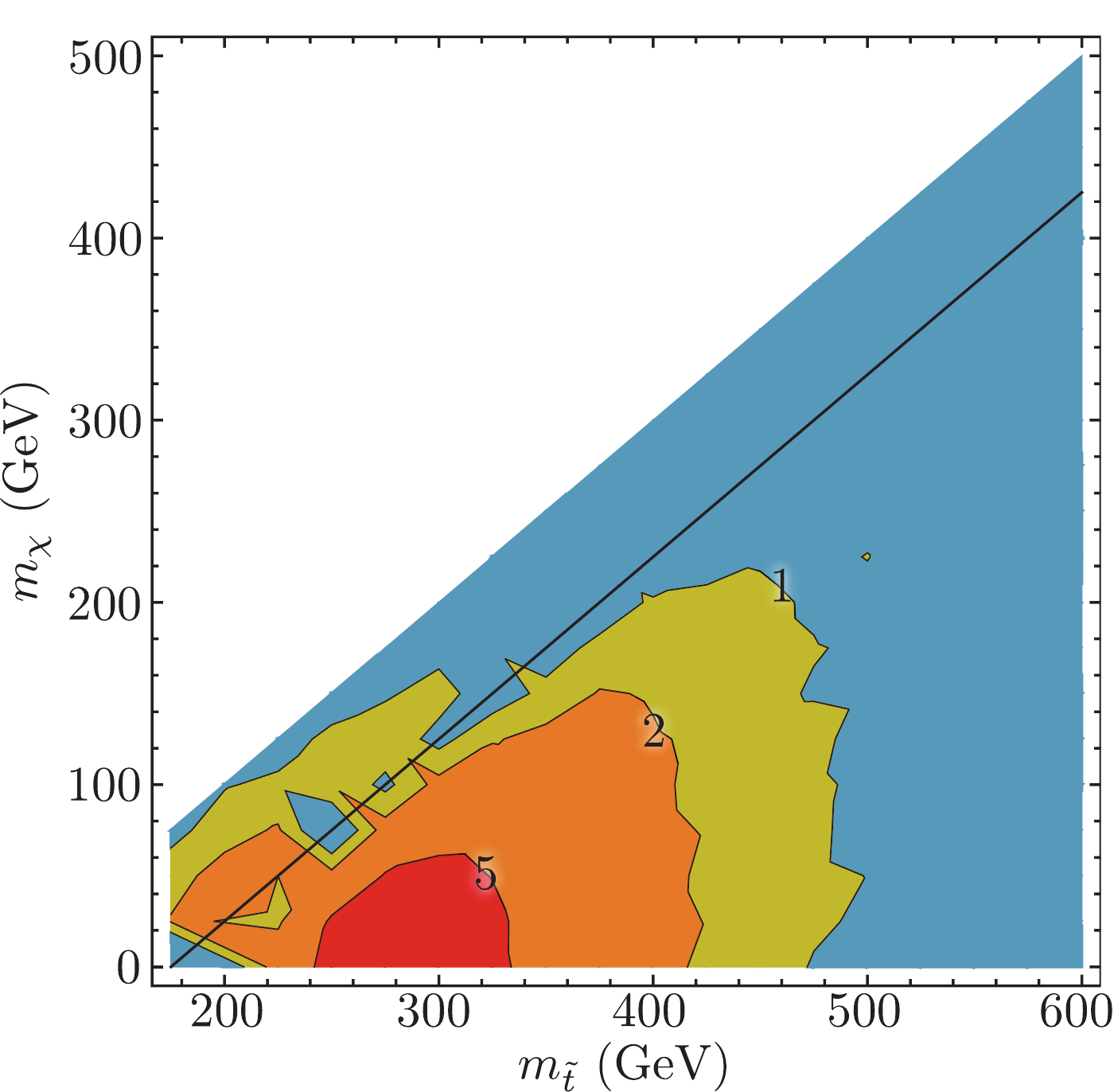}\includegraphics[width=0.4\columnwidth]{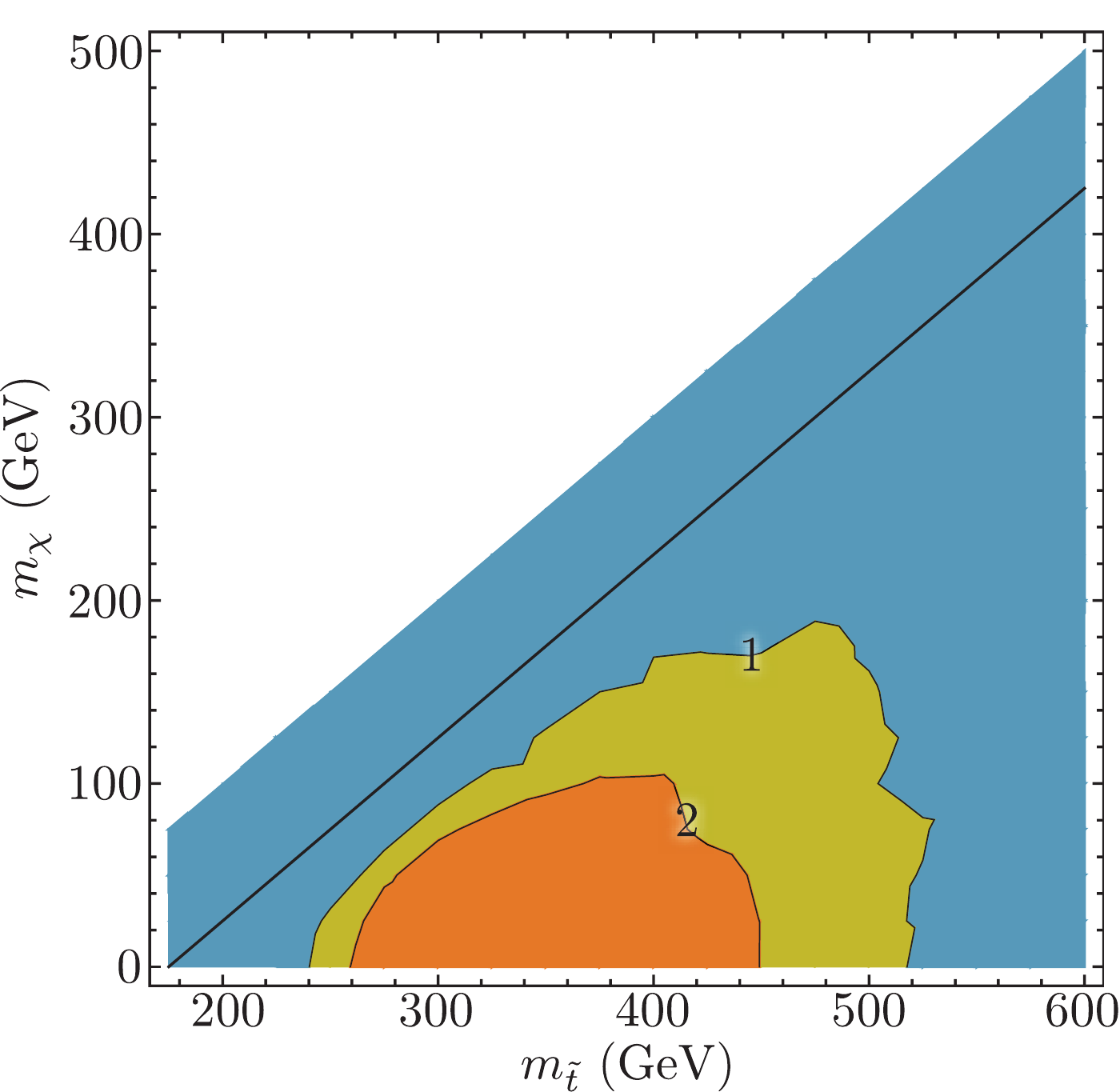}
\caption{Left: $S/\sqrt{O}$ for 150-250~GeV region of signal plus background $M_T^W$ distribution. Right: $S/\sqrt{O}$ for $250-400$~GeV region. These are computed with an unrealistic assumption of no systematics.\label{fig:MTWcc}}
\end{figure}

Using this fit and the associated errors, we repeat the profile likelihood analysis described previously, testing the background versus signal plus background hypotheses over 200 background-generated pseudo-experiments for each simplified model point. Our results are shown in Figure~\ref{fig:MTWexclusion}, for both the full profile likelihood analysis including all errors, and the case of errors inflated by a factor of 3.  The
sensitivity is similar to that obtained for the hadronic analysis.
 In Figure~\ref{fig:MTWcc}, we perform a cross-check using the cut-and-count method, with a background bin between $85-150$~GeV used for normalization, a low signal bin between $150-250$~GeV, and a high signal bin between $250-400$~GeV. As before, this simple analysis both validates and provides motivation for the full shape analysis.

\section{Conclusion \label{sec:conclusion}}

\begin{figure}[t]
\includegraphics[width=0.6\columnwidth]{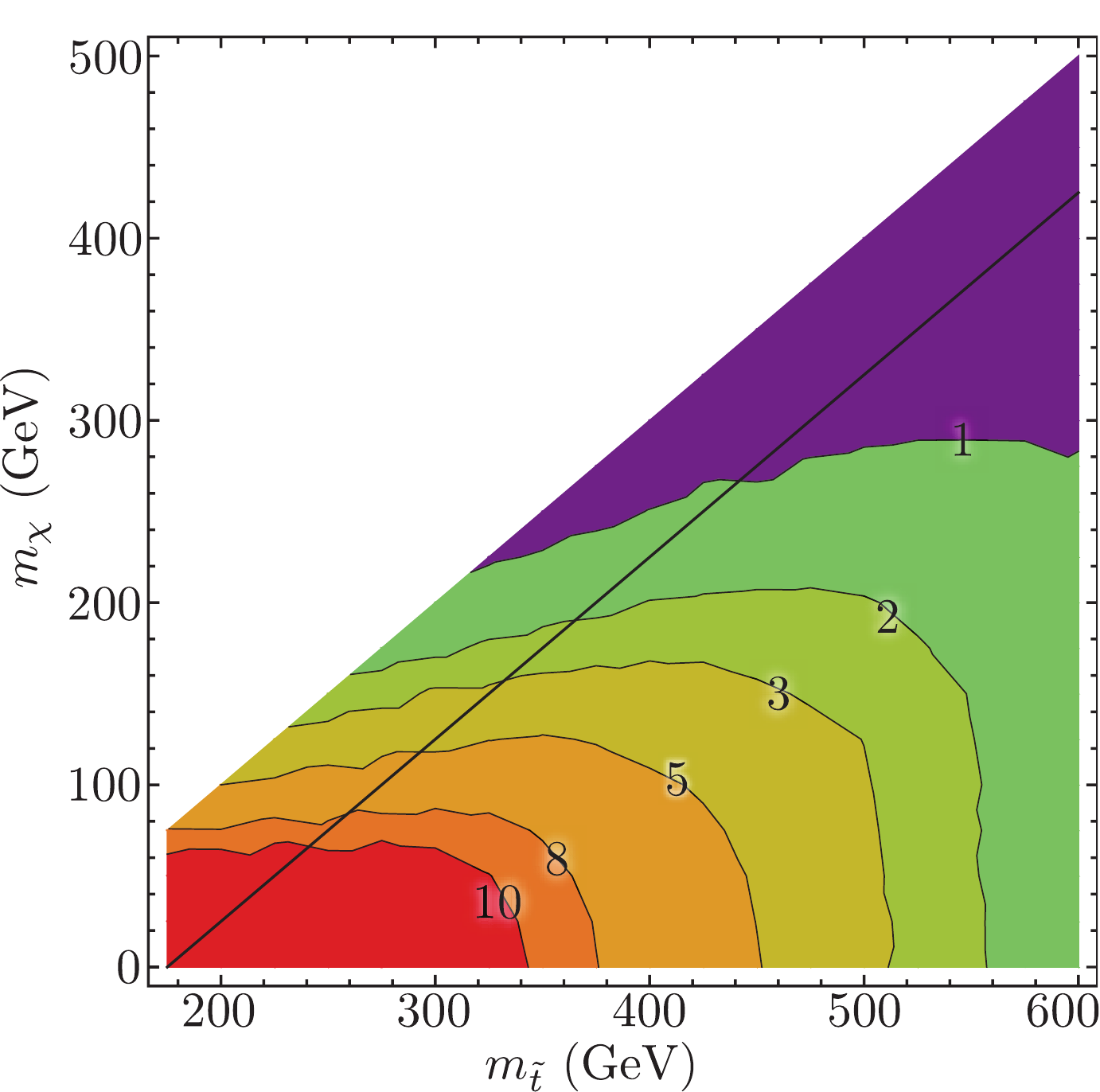}
\caption{Expected sensitivity, in standard deviations, to SUSY stop signals using a combination of $\met$ and $M_T^W$ shape analyses, included all fitting errors in the maximum likelihood method.}\label{fig:supercombo}
\end{figure}

Third generation squarks are an integral part of the supersymmetric solution to the naturalness and hierarchy problems. More generally, the large Yukawa couplings between the top and the Higgs hint at some connection between the third generation and electroweak symmetry breaking. Improving the search techniques for stop squarks (and more generally, top partners) at the LHC is therefore of great theoretical and experimental interest. In this paper, we have demonstrated that a dedicated search for stop pairs in hadronic and semi-leptonic channels has the potential to improve the current limits, especially for mass values such that the stop and the LSP + top quark system are nearly degenerate. 

We see that the tranverse momentum in the lab frame produced by the LSPs in stop pair decays is larger than naive expectations. Thus hadronic searches that limit the contribution to $\met$ from Standard Model neutrinos can provide significant discrimination between signal and background. The most obvious way to access this kinematic information is by modeling the shapes of $\met$ distributions for the most relevant SM backgrounds.  As we have shown, such an analysis is capable of excluding stops up to $\sim 250$~GeV in the degenerate case, as compared to up to $550$~GeV when the LSP is light.  However, we expect that other $\met$-based variables could also serve. For the semi-leptonic stop search, we saw that the most straightforward approach is to model the shape of the transverse mass variable $M_T^W$, which is related to $\met$. We found that the projected sensitivity to degenerate stops in the semi-leptonic case also reaches up to $\sim 300$~GeV, similar to that in the hadronic channel.  Finally, since these two channels are independent, we combine these bounds which we show in Figure~\ref{fig:supercombo}.  The resulting exclusion for light neutralinos is $560$ GeV and $360$ GeV in the degenerate case.

We note that the CMS Razor analyses \cite{Rogan:2010kb,PhysRevD.85.012004,CMSrazor,CMS-PAS-EXO-11-030} access the missing transverse momentum of an event through the transverse Razor variable $M_T^R$ (and through this, the Razor ratio $R$). As such, one would expect that Razor inclusive searches could be competitive with a more targeted analysis using the techniques outlined in this report. More generally, our $\met$ search could be upgraded to a multi-dimensional shape analysis as used in the Razor. 
Though, in this theoretical work, the analytic fits for the $\met$ distributions were drawn from Monte Carlo simulation,
the experimental collaborations can use data control samples to model the background shapes. In the real experimental
analyses the optimal baseline selections in both the hadronic and semi-leptonic channels could differ from those
presented here. Furthermore, we have shown that even if the extraction of the fit parameters from data suffers from considerably more uncertainty than our Monte Carlo based analysis the shape-based approach, unlike a cut and count, still has good reach. 

Our results support the assertion that it is not possible for stop squarks lighter than $\sim 1$ TeV in $R$-parity conserving
SUSY to elude LHC searches
over the long run. A stop discovery would be at least as fundamentally important as a Higgs discovery, while complete
exclusion of stops with mass lighter than a TeV would be a significant blow to our understanding of the connection
between supersymmetry and electroweak symmetry breaking.

%%%%%%%%%%%%%%%%%%%%%%%%%
\begin{acknowledgments}
The authors wish to thank Werner Porod, Maurizio Pierini, Will Reece, Chris Rogan, Maria Spiropulu, and Ciaran Williams for their advice and helpful discussions. MRB wishes to thank C.~Weinstein for suggesting the removal of the Appendix, which greatly expedited the completion of the paper. C.-T.Y is supported by the Fermilab Fellowship in Theoretical Physics. Fermilab is operated by Fermi Research Alliance, LLC, under contract DE-AC02-07CH11359 with the United States Department of Energy.
\end{acknowledgments}
%%%%%%%%%%%%%%%%%%%%%%%%%%
\bibliographystyle{apsrev}
\bibliography{invisiblerazor}

\end{document}